\documentclass[%
 reprint,
 amsmath,amssymb,
 aps,
]{revtex4-2}

\usepackage{CJKutf8}
\usepackage{graphicx}
\usepackage{dcolumn}
\usepackage{bm}
\usepackage{xcolor}

\begin{document}

\begin{CJK*}{GB}{} 

\preprint{APS/123-QED}

\title{Drift of a Charged Particle in a Black Hole Magnetosphere: \\ 
the Effect of Black Hole Spin}

\author{
\begin{CJK}{UTF8}{gbsn}
Ye Shen (沈冶)
\end{CJK}}
 \email{shenye199594@fzu.edu.cn}
\affiliation{%
College of Physics and Information Engineering, Fuzhou University, 
                    Fuzhou 350108, Fujian, P. R. China}%

\date{\today}

\begin{abstract}
Sparse plasma immersed in a strong magnetic field is common in astrophysical scenarios, for which a collisionless description is more appropriate. The guiding-center approximation serves as an effective tool for describing the motion of charged particles in a strong magnetic field, revealing the drift, a sub-leading order motion that causes particles to deviate from the magnetic field lines. It is widely accepted that the charged particles drift under the induction of gravity in Newtonian framework. However, a covariant formulation of the guiding-center approximation, as well as the drift of a charged particle induced by strong gravitational field, has yet to be fully explored. As a preliminary study, we aim to analyze the drift of a single charged particle in a black hole magnetosphere via a covariant approach. Particularly, we focus on the effects of black hole spin. We first examine the drift caused by the rotation of a massive object in the Newtonian regime using the language of gravitoelectromagnetism. A covariant formula of drift is derived, which is confirmed to reduce to the Newtonian results in the weak gravitational field and low speed limit. The impact of BH spin on the drift is analyzed in detail in Kerr spacetime specifically, for which we define the drift rate whose value quantifies the significance of the drift. Our results indicate that the impact of BH spin on the drift of a charged particle in a BH magnetosphere can be comparable to that of other factors in some extreme cases.
\end{abstract}

\maketitle

\end{CJK*}


\section{Introduction}
\label{sec:intro}

Evolution of plasma under the influence of a strong magnetic field is one of the most important topics in astrophysics. It helps explain many phenomena in different astrophysical environments, such as galaxy clusters, stellar evolution, and accretion flows around neutron stars or black holes (BHs). In general, this problem is simplified into two limiting cases: the limit that particles are fully collisional and the limit that particles are nearly collisionless. 

Magnetohydrodynamics (MHD) is the standard approach in the fully collisional limit, where the plasma is treated as a fluid \cite{Gammie:2003rj,DelZanna:2007pk,Stone:2008mh,Porth:2016rfi}. It is the most widely used numerical method for simulating accretion flows and relativistic jets \cite{Penna:2010,Tchekhovskoy:2011zx,Mishra:2019utv,Narayan:2021qfw,Liska:2022jdy,Hankla:2025jsl,Xia:2026nkt}. However, this collisional approximation breaks down in environments with sparse plasma and strong magnetic fields. Recent studies suggest that the mean free path of particles near supermassive BHs may exceed the size of the entire accretion system \cite{Frank2011IJMPD,EHT_2022,Luepker_2026}. Furthermore, MHD simulations indicate that the collisionless limit is more appropriate in the region $r\lesssim 10r_g$ (where $r_g$ is the gravitational radius of the BH) because of the high magnetization \cite{Chatterjee:2020eqc,Davelaar:2023dhl,Yang:2024kpz}. 

The particle-in-cell (PIC) model is a kinetic approach for describing nearly collisionless plasma, based on solving the Vlasov-Maxwell equations \cite{Haugbolle2013,Arber2015PPCF,Shalaby:2017yok}. It can capture microscale phenomena such as particle acceleration in magnetic reconnection \cite{Vega:2025afy,Gong:2025kxu,Bacchini:2025bbj,Hu2025}. However, due to the difficulty in solving the Vlasov-Maxwell equations, it is almost impossible to yield analytical solution under the PIC scheme. Moreover, its sensitivity to microscale physics restricts its ability to describe the large-scale evolution \cite{Parfrey:2018dnc,Crinquand:2020ppq,Galishnikova:2022mjg}.

\begin{figure}
    \centering
    \includegraphics[width=0.95\linewidth]{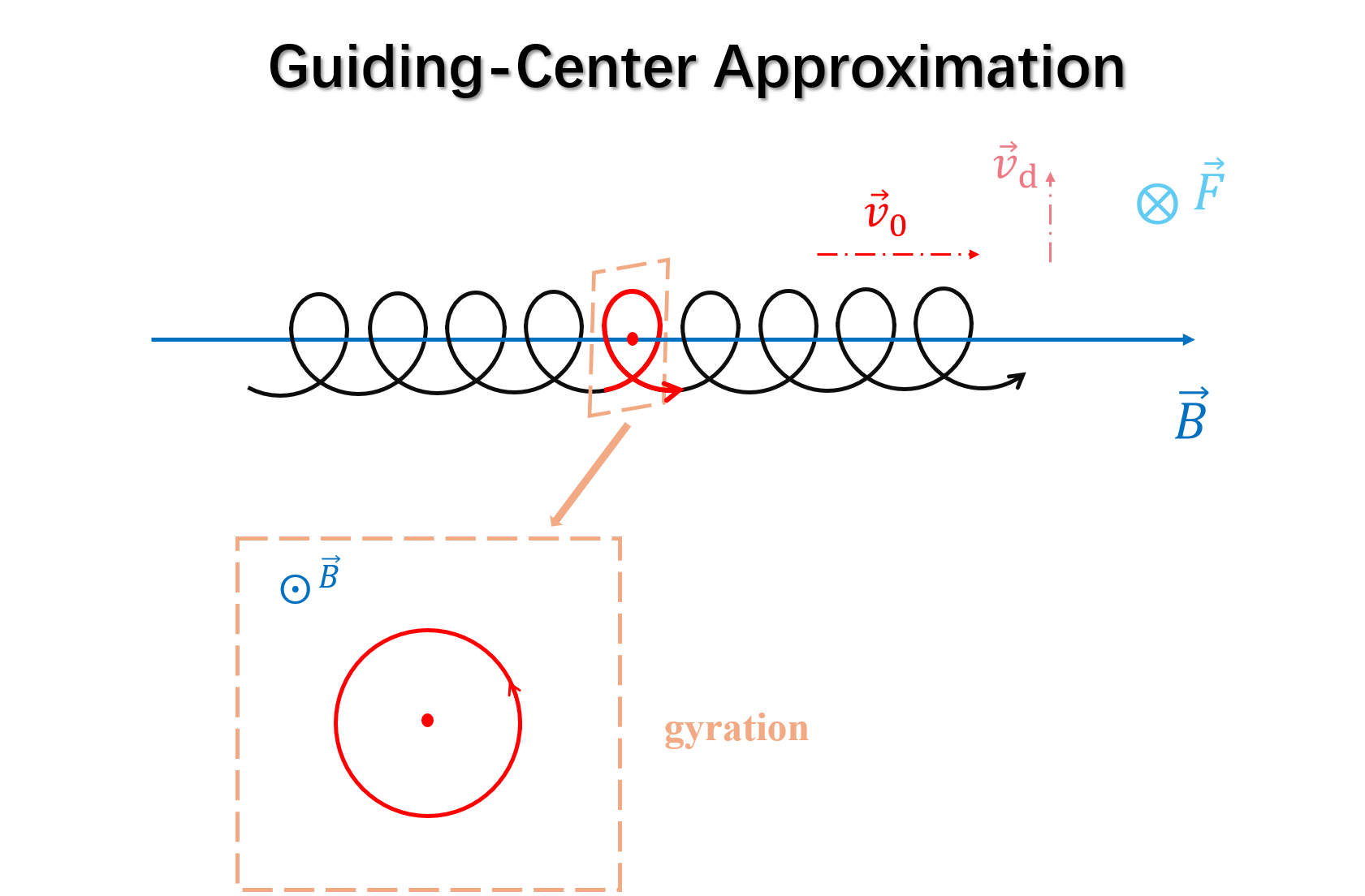}
    \caption{Schematic diagram of guiding-center approximation.}
    \label{fig:gca}
\end{figure}

The gyrokinetic (GK) model is another well-known kinetic approach, which is established upon the guiding-center approximation \cite{Beklemishev:2004qp,Howes:2005ms,Brizard2007RvMP,Numata:2010qr,Abel2013RPPh}. As we know, charged particles move in spiral orbits around strong magnetic field lines. In this framework, their motion could be decomposed into components parallel and perpendicular to the magnetic field (see Fig.~\ref{fig:gca}). The perpendicular part corresponds approximately to the periodic gyration. As long as the gyroradius is sufficiently small (as will be discussed in Sect.~\ref{sec:basic}), the trivial gyration can be integrated out (or ignored). The motion of the charged particles can then be approximated as the motion of the gyration center, called the guiding-center. Classically, the guiding-center would not move exactly along the magnetic field. Other forces acting on the charged particle (much weaker than the Lorentz force), the curvature of magnetic field lines and the gradient of magnetic strength make the guiding-center slowly deviate, a phenomenon known as drift. Knowledge of the drift is of vital importance for understanding the motion of charged particles under the guiding-center approximation.

The GK simulation is now well developed in the Newtonian framework, utilizing the guiding-center approximation and its higher-order extensions. However, its covariant formulation in the general relativistic (GR) scheme, along with the exact treatment of gravitational effects, has only been studied in detail in recent works. In Ref.~\cite{Beklemishev:2004}, a covariant formulation of the guiding-center approximation is derived using the tetrad formalism, but gravitational effects were hardly considered since the drift was evaluated in the local rest frame. Ref.~\cite{Trent:2024pii} covers almost all 1st order drifts known from classical theory. However, it only provides the equation of motion (EoM), rather than the covariant expression for drift velocity. In Ref.~\cite{Mizuno:2025mog}, a compact expression for the drift velocity is proposed by expanding the 4-velocity of the charged particle $u^{\mu}$ and the EM tensor $F_{\mu\nu}$ in a power series ordered by $m/e$, which is then applied to a two-fluid model previously discussed in flat spacetime \cite{Petrova:2014pea,Petrova:2016uuf}.

In this work, we aim to discuss the drift of a charged particle in the BH magnetosphere. More specifically, we focus on the effect of BH spin, which has not been systematically considered because it is much weaker than the drift induced by gravity $\Vec{g}$ in the Newtonian limit. However, the effect remains unclear when the charged particle is very close to the BH. As a preliminary study, we analyze only the motion of a single charged particle. We begin by examining the drift induced by the rotation of a compact object in the Newtonian framework. Then we apply the method proposed in Ref.~\cite{Mizuno:2025mog} to obtain the equations of motion (EoM) in the guiding-center approximation, from which we extract the information about the drift in the GR scheme. We will show that the formulae in GR reduce to the Newtonian results in the weak gravitational field and low speed limits. In two typical situations, we consider the importance of BH spin for the drift of a charged particle. We define the drift rate and the relative spin-induced drift rate to assess the significance of drift and the dominance of BH spin. Our main conclusions are as follows:
\begin{itemize}
    \item BH spin is the only factor causing the drift of a charged particle when the magnetic field is purely radial.
    \item The contribution of BH spin can become non-negligible for the drift, compared with those of BH mass and the curvature of magnetic field lines, when a charged particle moves extremely close to a rapidly rotating BH.
\end{itemize}

This manuscript is organized as follows. Basic assumptions are listed in Sect.~\ref{sec:basic}. In Sect.~\ref{sec:Newtonian}, we discuss the drift of a charged particle induced by a rotating massive object in the Newtonian scheme. We derive the covariant EoM under the guiding-center approximation in Sect.~\ref{sec:gca-GR} and confirm in Sect.~\ref{sec:limit} that the Newtonian formulation is recovered. The case in Kerr spacetime is considered in Sect.~\ref{sec:Kerr}, where the effect of BH spin and its nonrelativistic limit are presented. We demonstrate the significance of drift and the importance of BH spin in Sect.~\ref{sec:rate} by discussing the drift rate. Further discussions are made in Sect.~\ref{sec:discuss}. A brief summary is provided in Sect.~\ref{sec:sum}. In what follows, we adopt units with $G=c=1$, so that the units of length and time are equivalent to that of mass.

\section{Basic Settings}
\label{sec:basic}

We would like to discuss the drift, induced by BH spin, of a charged particle moving in the magnetosphere of this BH. Before proceeding with the calculations, let us summarize some basic assumptions and their consequences, which are appropriate for the schemes we consider.

{\bf 1.~Degenerate and Magnetically Dominated EM Field:}

The EM field considered in the following calculations is degenerate and magnetically dominated. In covariant language, an EM field $F_{\mu\nu}$ is degenerate if there exists a 4-vector $u^{\mu}$ satisfying:
\begin{equation}
    F_{\mu\nu}u^{\nu}=0
    \label{eq:degenerate}
\end{equation}
As long as $u^{\mu}$ is timelike, the field is magnetically dominated \cite{Gralla:2014yja,Shen:2026aye}. Physically, Eq.~\eqref{eq:degenerate} implies that there is no electric field in the rest frame of $u^{\mu}$. It can be shown that, for a degenerate EM field, $\Vec{E}$ is either zero or perpendicular to $\Vec{B}$ in an arbitrary rest frame.

A degenerate and magnetically dominated field is highly applicable in astrophysics. First, the resistivity (i.e., the magnetic diffusivity) in astrophysical plasma is mostly tiny. In this case, the ideal MHD becomes an effective description \cite{Frank2002}. In ideal MHD, the electric field vanishes in the rest frame of plasma, which preserves the degeneracy and magnetic dominance of the EM field. Second, whenever the EM stress-energy dominates (such as in the BH magnetosphere), force-free electrodynamics (FFE) becomes an effective description, in which the 4-current plays the role of timelike $u^{\mu}$ in Eq.~\eqref{eq:degenerate}. Therefore, in this work, it is appropriate to restrict our discussion to degenerate and magnetically dominated EM fields.

{\bf 2.~Guiding-Center Approximation:}

Guiding-center approximation is applicable to the motion of a charged particle only when the magnetic field is sufficiently strong. Specifically, the impact of the magnetic field on the charged particle must be much more significant than any other factors. Otherwise, the guiding-center approximation is not valid. Namely, the application of guiding-center approximation requires $F\ll eBv$, where $F$ is the magnitude of forces other than the Lorentz force acting on the charged particle \cite{Vandervoort1960,Northrop1961}. In this work, we analyze gravitational effects within guiding-center approximation and therefore presume that $mMr^{-2}\ll eBv$.

Meanwhile, the gyration of the charged particle around the magnetic field line is integrated out in the guiding-center approximation. For this purpose, the gyroradius $R_{\rm g}$ must be sufficiently small. Specifically, $R_{\rm g}$ should be much smaller than the scales characterizing the variations of magnetic field \cite{Vandervoort1960,Northrop1961}. There are two relevant characteristic scales. The first is the radius of curvature of the magnetic field line $\left|\Vec{\kappa}_B\right|^{-1}\equiv \left|\left(\hat{B}\cdot\nabla\right)\hat{B}\right|$, where $\hat{B}$ is the unit vector along the magnetic field. The second is the scale characterizing the variation of magnetic field strength $R_{\nabla B}:=\left|B^{-1}\nabla B\right|^{-1}$. In summary, the guiding-center approximation requires $R_{\rm g}\ll \left|\Vec{\kappa}_B\right|^{-1}$ and $R_{\rm g}\ll R_{\nabla B}$. Classically, it is expected that the gyration of a charged particle rapidly shrinks to a point in a strong magnetic field due to efficient synchrotron radiation. Therefore, a sufficiently small gyroradius is not a stringent assumption.

It should be noted that we consider the motion of only a single charged particle in this work. Hence the effect of this particle on the EM field is neglected. The degenerate EM field is previously determined as a background. It is different from Ref.~\cite{Mizuno:2025mog}, where a two-fluid model was considered so that the collective back-reaction of the plasma on the EM field is then non-negligible. We will reiterate this point in Sect.~\ref{sec:gca-Newtonian}.

\section{Newtonian Scheme}
\label{sec:Newtonian}

\subsection{Gravito-Electromagnetism}
\label{sec:gem}

Let us begin our discussion in the Newtonian scheme. Consider a charged particle moving under the influence of a rotating massive object and a surrounding strong magnetic field. The motion of the particle obeys the following equation \cite{Mashhoon:2003ax,Tartaglia:2003wx}:
\begin{equation}
    \frac{d\Vec{v}}{dt} = \Vec{g} + \Vec{v}\times\Vec{b} + \frac{e}{m}\Vec{v}\times\Vec{B}+\ldots
    \label{eq:gem-eom}
\end{equation}
where the ellipsis represents other types of forces. The third term on the right-hand side (RHS) is just the Lorentz force. The validity of Eq.~\eqref{eq:gem-eom} requires that the particle is nonrelativistic ($v\ll 1$), the size of the massive object is much smaller than its distance to the charged particle ($r\gg M$), and the angular momentum of the object is negligible compared with its mass ($J/M\ll M$). The quantities $\Vec{g}$ and $\Vec{b}$ on the RHS of Eq.~\eqref{eq:gem-eom} are referred to as the gravitoelectric and gravitomagnetic field strengths, respectively. Analogous to the treatment in electromagnetism, $\Vec{g}$ and $\Vec{b}$ could be derived from a scalar and a vector potential via \cite{Mashhoon:2003ax,Tartaglia:2003wx}:
\begin{equation}
    \Vec{g}=-\nabla\Phi-\partial_t\Vec{\mathcal{A}} ~~~,~~~
    \Vec{b}=\nabla\times\Vec{\mathcal{A}}
    \label{eq:gem-strength}
\end{equation}
When the gravitoelectromagnetic field is sourced by a rotating massive object, $\Phi$ and $\Vec{A}$ satisfy:
\begin{equation}
    \Phi = \frac{M}{r} ~~~,~~~
    \Vec{\mathcal{A}} = -\frac{2\Vec{J}\times\hat{r}}{r^2}
    \label{eq:gem-field}
\end{equation}
where $\Vec{r}$ is the vector pointing from the massive object to the charged particle. When the field is nearly stationary ($\partial_t \sim 0$), the gravitoelectric field reduces to the Newtonian gravity. The gravitomagnetic field, in contrast, arises from the rotation of the massive object. 

Specifically, for a massive object with constant mass $M$ and angular momentum $\Vec{J}=J\hat{k}$, the field strengths at position $\Vec{r}=x\hat{i}+y\hat{j}+z\hat{k}$ (with the massive object placed at the origin) are given by:
\begin{equation}
    \Vec{g}=-\frac{M}{r^2}\hat{r} ~~,~~
    \Vec{b}=-\frac{2J}{r^5}\left[3xz\hat{i}+3yz\hat{j}-\left(r^2-3z^2\right)\hat{k}\right]
    \label{eq:gem-field2}
\end{equation}
On the equatorial plane (the plane perpendicular to $\Vec{J}$), the gravitomagnetic field reduces approximately to:
\begin{equation}
    \Vec{b}\approx\frac{2J}{r^3}\hat{k}
    \label{eq:gem-field3}
\end{equation}
It is clear that the gravitomagnetic effect is much weaker than the gravitoelectric one under the limit of $r\gg M$.

\subsection{Guiding-Center Approximation}
\label{sec:gca-Newtonian}

As stated in Sect.~\ref{sec:basic}, the guiding-center approximation is effective for describing the motion of a charged particle when magnetic effects dominate. In the guiding-center approximation, the gyration of the particle is integrated out. Instead of tracking the exact trajectory of the charged particle, we only need to determine the trajectory of the gyration center, known as the guiding-center. Generally, the velocity of the guiding-center could be expanded as:
\begin{equation}
    \Vec{v} = \Vec{v}_0 + \Vec{v}_1 + \ldots
    \label{eq:gca-Newtonian}
\end{equation}
The 0th order term gives a constant velocity along the magnetic field line, which, together with the gyration, describes a spiral motion of the charged particle along the magnetic field. The 1st order term corresponds to the drift, a perpendicular migration of the guiding-center relative to the magnetic field. When a force $\Vec{F}$, much weaker than the Lorentz force, acts on the charged particle, the drift velocity of the guiding-center obeys:
\begin{equation}
    \Vec{v}_1\equiv\Vec{v}_{\rm d} = \frac{\Vec{F}\times\hat{B}}{eB}
    \label{eq:vd-Newtonian}
\end{equation}
The drift can also be induced by magnetic curvature and the gradient of magnetic field strength. These two factors will also be discussed later. When $\Vec{F}$ is generated by a rotating massive object, based on Eq.~\eqref{eq:gem-eom}, we have:
\begin{equation}
    \Vec{F} \approx m\Vec{g} + m\Vec{v}_0 \times \Vec{b} = m\Vec{g} + m v_0\hat{B}\times\Vec{b}
    \label{eq:gem-force}
\end{equation}
where we have used $\Vec{v}\approx\Vec{v}_0$ in the first step and $\Vec{v}_0=v_0\hat{B}$ in the second. Substituting Eq.~\eqref{eq:gem-force} into Eq.~\eqref{eq:vd-Newtonian} yields:
\begin{equation}
    \Vec{v}_{\rm d} = \frac{m}{eB}\Vec{g}\times\hat{B}
    +\frac{mv_0}{eB}\Vec{b}_{\perp}
    \label{eq:vd-Newtonian2}
\end{equation}
where  $\Vec{b}_{\perp}:=\left[\Vec{b}-\left(\hat{B}\cdot\Vec{b}\right)\hat{B}\right]$ is the component of the gravitomagnetic field perpendicular to the magnetic field. The 1st term on the RHS of Eq.~\eqref{eq:vd-Newtonian2} is just the widely known drift induced by gravity, while the 2nd term is the drift induced by the gravitomagnetic force arising from the rotation of the massive object.

\begin{figure}
    \centering
    \includegraphics[width=\linewidth]{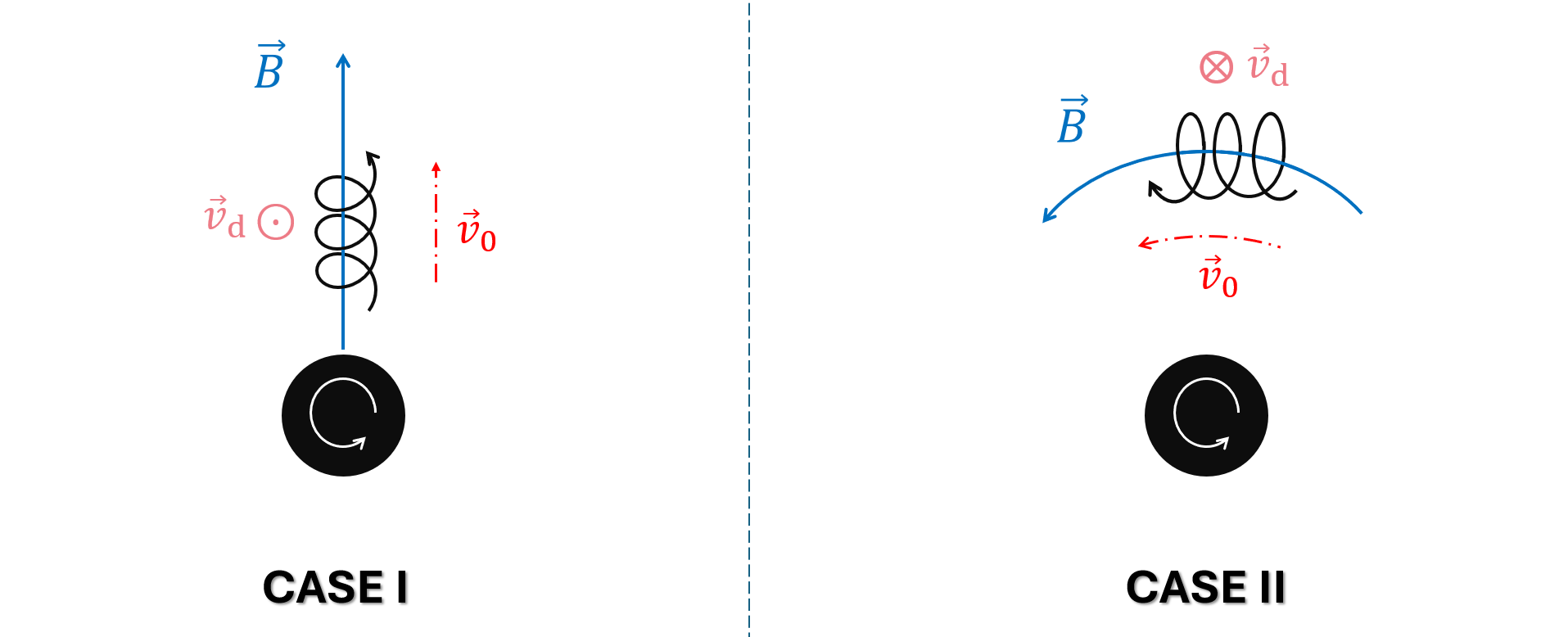}
    \caption{Schematic diagram of \textit{case I} (left panel) and \textit{case II} (right panel).}
    \label{fig:cases}
\end{figure}

As a preliminary study, let us assume that the angular momentum of the massive object is constant along $\hat{k}$, while the charged particle moves nearly on the equatorial plane, meaning that $\Vec{v}_0$ and $\hat{B}$ have no component along $\hat{k}$. Eq.~\eqref{eq:vd-Newtonian2} then becomes:
\begin{equation}
    \Vec{v}_{\rm d} = -\frac{mM}{eB}\frac{\hat{r}\times\hat{B}}{r^2} + \frac{2mv_0J}{eB}\frac{\hat{k}}{r^3}
    \label{eq:vd-Newtonian3}
\end{equation}
Two specific cases are worth discussing. \textit{Case I}: the magnetic field is purely radial on the equatorial plane (left panel of Fig.~\ref{fig:cases}). In this situation, the 1st term on the RHS vanishes. The rotation of the massive object is the only effect that causes the charged particle to drift away from the equatorial plane. \textit{Case II}: the magnetic field is purely toroidal on the equatorial plane (right panel of Fig.~\ref{fig:cases}). In this case, both the gravitoelectric and gravitomagnetic fields cause the charged particle to drift away from the equatorial plane. It is clear from Eq.~\eqref{eq:vd-Newtonian3} that the drift induced by gravity (or gravitoelectricity) dominates in the limit $r\gg M$. In other words, within the Newtonian scheme, the drift of a charged particle induced by the rotation of a massive object is largely negligible.

\section{General Relativistic Scheme}
\label{sec:GR}

\subsection{Guiding-Center Approximation in Covariant Formulation}
\label{sec:gca-GR}

Now let us discuss the guiding-center approximation in the GR scheme. The covariant EoM for a charged particle is:
\begin{equation}
    \iota u^{\mu}\nabla_{\mu}u^{\nu} = F^{\nu}_{~\kappa}u^{\kappa} + f^{\nu}/e
    \label{eq:GR-eom}
\end{equation}
Here $F^{\nu}_{~\kappa}$ is the electromagnetic tensor, so that $F^{\nu}_{~\kappa}u^{\kappa}$ includes both the Coulomb and Lorentz force, while $f^{\nu}$ is the 4-vector of other forces. We denote $\iota=m/e$ temporarily. When the guiding-center approximation is applicable, the 4-velocity of a charged particle can be similarly expanded as:
\begin{equation}
    u^{\mu} = u_0^{\mu} + u_1^{\mu} + u_2^{\mu} + \ldots
    \label{eq:gca-GR}
\end{equation}
Based on the Newtonian results, it is expected that $n$-th order term in Eq.~\eqref{eq:gca-GR} would be proportional to $\iota^n$. Accordingly, we rewrite Eq.~\eqref{eq:gca-GR} in the form of \cite{Mizuno:2025mog}:
\begin{equation}
    u^{\mu} = u_0^{\mu} + \iota\left(\iota^{-1}u_1^{\mu}\right) + \iota^2\left(\iota^{-2}u_2^{\mu}\right) + ...
    \label{eq:gca-GR2}
\end{equation}
For infinitesimal $\iota$ (which makes the guiding-center approximation valid), the factor $\iota^{-n}u_n^{\mu}$ remains finite. Similarly, since the Lorentz force dominates, we at least have:
\begin{equation}
    f^{\mu} = 0 + \iota\left(\iota^{-1}f_1^{\mu}\right) + \iota^2\left(\iota^{-2}f_2^{\mu}\right) + ...
    \label{eq:gca-f}
\end{equation}
Substituting Eq.~\eqref{eq:gca-GR2} and \eqref{eq:gca-f} into Eq.~\eqref{eq:GR-eom}, one could derive the EoM at different orders of $\iota$ like \cite{Mizuno:2025mog}:
\begin{equation}
    \begin{aligned}
        \mathcal{O}\left(\iota^0\right):&~~
        eF^{\nu}_{~\kappa}u_0^{\kappa} = 0 \\
        \mathcal{O}\left(\iota^1\right):&~~
        eF^{\nu}_{~\kappa}u_1^{\kappa} = m u_0^{\mu}\nabla_{\mu}u_0^{\nu} - f_1^{\nu} \\
        \mathcal{O}\left(\iota^2\right):&~~
        eF^{\nu}_{~\kappa}u_2^{\kappa} = m u_0^{\mu}\nabla_{\mu}u_1^{\nu} + m u_1^{\mu}\nabla_{\mu}u_0^{\nu} - f_2^{\nu} \\
    \end{aligned}
    \label{eq:GR-eom2}
\end{equation}
The 0th order equation implies that the guiding-center is stuck on the EM field. Comparing with Eq.~\eqref{eq:degenerate}, this actually preserves the degeneracy and magnetic dominance of the EM field. The 1st order EoM is expected to provide the information about the drift, as $u_1^{\mu}$ should be proportional to $\iota\equiv m/e$. Once the spacetime and $F_{\mu\nu}$ are specified, one could uniquely determine $u_0^{\mu}$ from the 0th order equation in Eq.~\eqref{eq:GR-eom2}, $u_1^{\mu}$ via the 1st order equation, and so on.

It is worth reiterating here that the EM field is not expanded in series as: $F=F_0^{\mu\nu}+\iota\left(\iota^{-1}F_1^{\mu\nu}\right)+\iota^2\left(\iota^{-2}F_2^{\mu\nu}\right)+ ...$ in this work, different from the treatment in Ref.~\cite{Mizuno:2025mog}. In Ref.~\cite{Mizuno:2025mog}, a two-fluid model of plasma was taken into consideration. The collective back-reaction of the plasma is non-negligible, which contributes to the higher order terms of the field. The field is prescribed as a background in this work, where the back-reaction of the single particle is neglected. In this sense, a non-zero $F_1^{\mu\nu}$ corresponds to a small electric field in the rest frame of $u_0^{\mu}$, breaking the degeneracy of the total EM field. Hence, assuming $F_n^{\mu\nu}=0$ is actually equivalent (up to $\mathcal{O}(m/e)$ at least) to our basic assumption that the prescribed EM field is exactly degenerate.

In general, the EM field can be fully decomposed into its electric and magnetic parts as: \cite{Uchida1997-1,Uchida1997-2,Gralla:2014yja}:
\begin{equation}
    F_{\mu\nu} = -\mathcal{E} \epsilon_{\mu\nu} + \mathcal{B} \epsilon^{\perp}_{\mu\nu}
    \label{eq:EB}
\end{equation}
where $\mathcal{E}^2$ and $\mathcal{B}^2$ represent the total electric and magnetic energy, respectively. Note that $\mathcal{E}$ and $\mathcal{B}$ are not the electric and magnetic field strengths observed in a specific rest frame, but the identity $E^2-B^2=\mathcal{E}^2-\mathcal{B}^2$ always holds. Indeed, $\pm \mathcal{E}$ and $\pm \mathcal{B}$ are the four eigenvalues of the $4\times 4$ matrix $\left[F_{\mu\nu}\right]$. The tensors $\epsilon_{\mu\nu}$ and $\epsilon^{\perp}_{\mu\nu}$ are two antisymmetric tensors, which could be regarded as the proper volume forms of the electric and magnetic sectors. Their normalizations are given by:
\begin{equation}
    -\epsilon_{\mu\nu}\epsilon^{\mu\nu}
    =\epsilon^{\perp}_{\mu\nu}\epsilon_{\perp}^{\mu\nu}=2 ~~~,~~~
    \epsilon_{\mu\kappa}\epsilon_{\perp}^{\kappa\nu}=0
    \label{eq:epsilon}
\end{equation}
The first two equations are basic properties of proper volume forms for 2D manifolds, indicating that the electric and magnetic field sheets are Lorentzian and Riemannian, respectively. With $\epsilon_{\mu\nu}$ and $\epsilon^{\perp}_{\mu\nu}$, the metric tensor of the spacetime can be decomposed as:
\begin{equation}
    g_{\mu\nu}=\epsilon_{\mu\kappa}\epsilon^{\kappa}_{~\nu}-\epsilon^{\perp}_{\mu\kappa}\epsilon^{\perp\kappa}_{~~~\nu}
    \label{eq:g-epsilon}
\end{equation}
implying that the entire 4D manifold could be decomposed into the 2D electric and magnetic sectors. 

A degenerate and magnetically dominated field satisfies $\mathcal{E}=0$ \cite{Uchida1997-1,Uchida1997-2,Gralla:2014yja}. In this case, we have:
\begin{equation}
    F_{\mu\nu}= \mathcal{B}\epsilon^{\perp}_{\mu\nu}
    ~~\Longleftrightarrow~~
    \epsilon^{\perp}_{\mu\nu}= \frac{1}{\mathcal{B}}F_{\mu\nu}
    \label{eq:F-degenerate}
\end{equation}
The 0th order EoM, together with Eq.~\eqref{eq:F-degenerate}, indicates that $u_0^{\mu}$ is stuck on the magnetic sector. Thus, using Eq.~\eqref{eq:g-epsilon}, we obtain:
\begin{equation}
    u^0_{\mu}\equiv g_{\mu\nu}u_0^{\nu}
    =\epsilon_{\mu\kappa}\epsilon^{\kappa}_{~\nu}u_0^{\nu}
\end{equation}
meaning that the tensor $\epsilon_{\mu\kappa}\epsilon^{\kappa}_{~\nu}$ acts as the metric tensor for the vector stuck on the magnetic sector. With Eq.~\eqref{eq:g-epsilon}, it is feasible to decompose $u_1^{\mu}$ into $u_{1,\perp}^{\mu}$ and $u_{1,\parallel}^{\mu}$, satisfying:
\begin{equation}
    0=\epsilon^{\mu}_{~\nu}u_{1,\perp}^{\nu}=\epsilon^{\perp\mu}_{~~~\nu}u_{1,\parallel}^{\nu}
    \label{eq:2-u1}
\end{equation}
with which $\epsilon^{\perp}_{\mu\kappa}\epsilon^{\perp\kappa}_{~~~\nu}$ and $\epsilon_{\mu\kappa}\epsilon^{\kappa}_{~\nu}$ play the role of metric tensors for $u_{1,\perp}^{\mu}$ and $u_{1,\parallel}^{\mu}$, respectively. Thus, contracting $\epsilon^{\perp}_{\mu\nu}$ on both sides of the 1st order EoM yields:
\begin{equation}
    \begin{aligned}
        u_{1,\perp}^{\mu}
        &=\frac{m}{e\mathcal{B}}\epsilon_{\perp}^{\mu\kappa}u_0^{\nu}\nabla_{\nu}u^0_{\kappa} - \frac{1}{e\mathcal{B}}\epsilon^{\mu\nu}_{\perp}f^1_{\nu} \\
        &=-\frac{m}{e\mathcal{B}}u_0^{\kappa}u_0^{\nu}\nabla_{\nu}\epsilon^{\perp\mu}_{~~~\kappa} - \frac{1}{e\mathcal{B}}\epsilon^{\perp\mu}_{~~~\nu}f_1^{\nu}
    \end{aligned}
    \label{eq:u1-1}
\end{equation}
This gives the covariant expression for the sub-leading order motion of guiding-center perpendicular to the magnetic sector, and it is expected to contain the information about the drift \cite{Mizuno:2025mog}. In the second equality of Eq.~\eqref{eq:u1-1} we have used $u_0^{\kappa}\epsilon^{\perp\mu}_{~~~\kappa}=0$. A more explicit expression reads:
\begin{equation}
    u_{1,\perp}^{\mu} = -\frac{m}{e\mathcal{B}}u_0^{\kappa}u_0^{\nu}\left(
    \partial_{\nu}\epsilon^{\perp\mu}_{~~~\kappa}
    -\Gamma^{\lambda}_{\nu\kappa}\epsilon^{\perp\mu}_{~~~\lambda}
    \right) 
    - \frac{1}{e\mathcal{B}}\epsilon^{\perp\mu}_{~~~\nu}f_1^{\nu}
    \label{eq:u1-2}
\end{equation}
where $\Gamma^{\lambda}_{\nu\kappa}$ are the Christoffel symbols. As we will see, the drift of the charged particle induced by gravity and the gravitomagnetic field arises from the 2nd term on the RHS of Eq.~\eqref{eq:u1-2}. Additionally, the drift caused by the curvature of magnetic field lines is contained either in the 1st or 2nd term, depending on the choice of coordinates. The drift induced by other forces comes from the 3rd term.

\subsection{Weak gravitational field and Low Speed Limits}
\label{sec:limit}

Let us now apply Eq.~\eqref{eq:GR-eom2} and Eq.~\eqref{eq:u1-2} to the weak gravitational field and low speed limit to see how the drifts induced by different factors in classical theory arise.

Assume that the motion of a charged particle is observed in some rest frame, and its 3-velocity is $\Vec{v}=\Vec{v}_0+\Vec{v}_1+\ldots$, with $\Vec{v}_i$ corresponding to $u^i$. The electric and magnetic field strengths are $\Vec{E}$ and $\Vec{B}$ in this rest frame, with $E$ and $B$ denoting their magnitude, respectively. The 0th order EoM in Eq.~\eqref{eq:GR-eom2} implies $\Vec{E}+\Vec{v}_0\times\Vec{B}=0$, from which we have:
\begin{equation}
    \Vec{v}_0=\left(\Vec{v}_0\cdot\hat{B}\right)\hat{B}+\frac{\Vec{E}\times\hat{B}}{B}
    \label{eq:v0-lim}
\end{equation}
Here, the 1st term is the velocity along the magnetic field, while the 2nd term is the drift induced by the electric field in the classical guiding-center approximation. Hence, from the GR perspective, the electric field induced drift comes from the 0th order EoM. It is actually a result of coordinate transformation. Alternatively, the drift velocity under the induction of the electric field is independent of $\iota\equiv m/e$, and therefore can only appear in the zeroth-order EoM.

The metric in a weak gravitational field could be expressed as $g_{\mu\nu}=\eta_{\mu\nu}+h_{\mu\nu}$, where $\eta_{\mu\nu}$ is the flat spacetime metric and $h_{\mu\nu}$ is the perturbation. We choose the Cartesian coordinates so that $\eta_{\mu\nu}={\rm diag}(-1,1,1,1)$ and $h_{\mu\nu}\ll 1$. It is convenient to set $h_{00}=-2\Phi$ and $h_{0i}=h_{i0}=\mathcal{A}_i$. The geodesic equations in the low speed limit yield $\Vec{g}=-\nabla\Phi-\partial_t\Vec{\mathcal{A}}$ and $\Vec{b}=\nabla\times\Vec{\mathcal{A}}$, giving rise to the theory of gravitoelectromagnetism. The nonvanishing components of the Christoffel symbols are:
\begin{equation}
    \begin{aligned}
        & \Gamma^0_{i0}=-\nabla_i\Phi ,~~~
        \Gamma^i_{00}=\nabla_i\Phi+\partial_t \mathcal{A}_i=-g_i , \\
        & \Gamma^j_{i0}=\left(\partial_i \mathcal{A}_j-\partial_j \mathcal{A}_i\right)=\varepsilon_{ijk}b_k , \\
        & \Gamma^0_{ij}=-\frac{1}{2}\left(\partial_i \mathcal{A}_j+\partial_j \mathcal{A}_i\right) , \\
        & \Gamma^k_{ij}=-\delta_{ij}\nabla_k\Phi+\delta_{ki}\nabla_m\Phi+\delta_{km}\nabla_i\Phi
    \end{aligned}
    \label{eq:Gamma}
\end{equation} 
with $\delta_{ij}$ the Kronecker delta and $\varepsilon_{ijk}$ the 3D Levi-Civita tensor. 

In the low speed limit $v_0\ll 1$, we have $u_0^{\mu}\approx \left(1,\Vec{v}_0\right)$ and consequently $E\ll B$, $\mathcal{B}\approx B$. Classically, we are interested in the situation where the electric field induced drift can be regarded as a perturbation. Namely, we require $E/B\ll v_0$ and hence $\Vec{v}_0\approx v_0\hat{B}$. Substituting all these into Eq.~\eqref{eq:u1-2}, one obtains:
\begin{equation}
    \begin{aligned}
        v_d^i \approx u_{1,\perp}^{i} = \frac{m}{eB} &\left\{\varepsilon_{ijm}v_0^2\hat{B}_{j}\hat{B}_{k}\partial_k\hat{B}_m \right. \\
        & + \varepsilon_{ijm}g_j \hat{B}_m - \varepsilon_{ijm}\varepsilon_{jkn}b_kv_{0,n}\hat{B}_m \\
        & + \frac{1}{m}\varepsilon_{ijm}\hat{B}_mf_{1,j} \\
        & \left. + \mathcal{O}\left(\Gamma v_0^2\right) + \mathcal{O}\left(Ev_0/B\right) + \mathcal{O}\left(Ef_1/B\right) \right\} 
    \end{aligned}
    \label{eq:v1-lim}
\end{equation}
where we have also used $\epsilon^{\perp i}_{~~~0}\approx -E_i/B$ and $\epsilon^{\perp i}_{~~~j}\approx -\varepsilon_{ijm}\hat{B}_m$. The 1st, 2nd and 3rd lines on the RHS of Eq.~\eqref{eq:v1-lim} come from the 1st, 2nd and 3rd terms on the RHS of Eq.~\eqref{eq:u1-2}, respectively.

It is evident that the two terms on the 2nd line of Eq.~\eqref{eq:v1-lim} are $\Vec{g}\times\hat{B}$ and $\Vec{v}_0\times \Vec{b}\times \hat{B}$, based on the definition of cross product. Therefore, the 2nd term on the RHS of Eq.~\eqref{eq:u1-2} turns out to be the drift induced by the gravitoelectromagnetic field. The 1st term on the 2nd line is the gravity induced drift, and the 2nd one is the drift induced by gravitomagnetism, namely by the rotation of the massive object. The GR results in the weak gravitational field and low speed limits are identical to the Newtonian results.

The 1st line of Eq.~\eqref{eq:v1-lim} contributes:
\begin{equation}
    \Vec{v}_{\Vec{\kappa}_B}=\frac{mv_0^2}{eB}\hat{B}\times\left(\hat{B}\cdot\nabla\right)\hat{B}
    \equiv \frac{mv_0^2}{eB}\hat{B}\times \Vec{\kappa}_B
    \label{eq:vd-curvature}
\end{equation}
where $\Vec{\kappa}_B\equiv \left(\hat{B}\cdot\nabla\right)\hat{B}$ is the curvature of the magnetic field line. Hence, the drift induced by the curvature of magnetic field line arises from the 1st term on the RHS of Eq.~\eqref{eq:u1-2}. It is worth noting that we have chosen the Cartesian coordinates, so all axes are straight. If we choose a coordinate system whose spatial axes are not straight (such as spherical coordinates), the curvature induced drift would arise from the 2nd term of Eq.~\eqref{eq:u1-2}. The 3rd line contributes $f\times\hat{B}$, which is identical to Eq.~\eqref{eq:vd-Newtonian}, representing the drift induced by other forces. The 4th line contains higher-order perturbations, which are neglected in the weak gravitational field and low speed limits and therefore do not appear in the Newtonian scheme.

The drift induced by the gradient of magnetic field strength does not appear. In the classical guiding-center approximation, the gradient of magnetic field strength makes a charged particle drift with velocity:
\begin{equation}
    \Vec{v}_{\nabla B} = \frac{m v_{\perp}^2}{2eB^2}\hat{B}\times\nabla B
    \label{eq:vd-gradient}
\end{equation}
where $v_{\perp}$ is the gyration speed of the charged particle perpendicular to $\Vec{B}$. The disappearance of $\Vec{v}_{\nabla B}$ in Eq.~\eqref{eq:u1-2} is because the information about gyration is not contained in power series of $u^{\mu}$. It is a drawback of the method introduced in Sect.~\ref{sec:gca-GR} and we will mention this point again in Sect.~\ref{sec:discuss}.

In summary, Eq.~\eqref{eq:u1-2} gives the drift velocity of a charged particle induced by almost all relevant factors.

\subsection{Kerr Spacetime}
\label{sec:Kerr}

Now let us specialize the problem to Kerr spacetime, the spacetime with a spinning BH at the origin. The line element in the Boyer-Lindquist (BL) coordinates reads:
\begin{equation}
    \begin{aligned}
        ds^2 = &-\left(1-\frac{2Mr}{\Sigma}\right)dt^2 
        + \frac{\Sigma}{\Delta}dr^2 + \Sigma d\theta^2 \\
        & + \frac{A\sin^2\theta}{\Sigma} d\phi^2 
        - \frac{4Mar\sin^2\theta}{\Sigma} dtd\phi
    \end{aligned}
    \label{eq:Kerr}
\end{equation}
with:
\begin{equation}
    \begin{aligned}
        \Sigma &= r^2+a^2\cos^2\theta~~,~~
        \Delta = r^2-2Mr+a^2 ~~, \\
        A &= \left(r^2+a^2\right)^2-a^2\Delta\sin^2\theta
    \end{aligned}
    \label{eq:sigma-delta}
\end{equation}
Here $a:=J/M$ is the spin parameter ranging from -1 to 1. We focus only on the charged particle moving on the equatorial plane ($\theta=\pi/2$). The nonvanishing components of the Christoffel symbols are listed in Appen.~\ref{sec:Christoffel}.

In accordance with the discussions in Sect.~\ref{sec:gca-Newtonian}, let us consider the two cases shown in Fig.~\ref{fig:cases}. \textit{Case I} assumes that the magnetic field lies radially on the equatorial plane, for which we have:
\begin{equation}
    \epsilon^{\perp}=\sqrt{\frac{\Sigma^2\Delta}{r^2-2Mr}}~d\theta\wedge d\phi
    \label{eq:epsilon-case1}
\end{equation}
Here we use the notation of differential form, with $\wedge$ representing the wedge product. In this case, only the $t$ and $r$ components of $u_0^{\mu}$ are nonvanishing, confirming that $\epsilon^{\perp}_{\mu\nu}u_0^{\nu}\equiv 0$. Thus, to 0th order, the guiding-center moves radially in BL coordinates. The 1st term on the RHS of Eq.~\eqref{eq:u1-2} then vanishes. The only nonvanishing component of $u_{1,\perp}^{\mu}$, coming from the 2nd term of Eq.~\eqref{eq:u1-2}, is the $\theta$ component (note that both $u^{\theta}$ and $u^{\phi}$ are angular velocities in BL coordinates, with units of $M^{-1}$):
\begin{equation}
    u_{1,\perp}^{\theta} = \frac{2mJ}{e\mathcal{B}} \frac{u_0^t u_0^r}{r^2\sqrt{\Delta^2-a^2\Delta}}
    \label{eq:u1-case1}
\end{equation}
This indicates that the charged particle drifts away from the equatorial plane under the induction of BH spin. In the limit $r\gg M$ and $u_0^t\rightarrow 1$, Eq.~\eqref{eq:u1-case1} reduces to:
\begin{equation}
    \Vec{v}_{\rm d} \approx ru_{1,\perp}^{\theta} \rightarrow \frac{2mv_0 J}{eB}\frac{\hat{\theta}}{r^3}
\end{equation}
which also reproduces the Newtonian results discussed in the paragraph following Eq.~\eqref{eq:vd-Newtonian3}.

In \textit{case II}, the magnetic field is purely toroidal on the equatorial plane, for which we have:
\begin{equation}
    \epsilon^{\perp}=\frac{\Sigma}{\sqrt{\Delta}}dr\wedge d\theta
    \label{eq:epsilon-case2}
\end{equation}
Now the only nonvanishing components of $u_0^{\mu}$ are $t$ and $\phi$ components. The 1st term on the RHS of Eq.~\eqref{eq:u1-2} vanishes again, while the 2nd term yields:
\begin{equation}
    u_{1,\perp}^{\theta}=-\frac{mM}{eB}\frac{\sqrt{\Delta}}{r^4}
    \left[u_0^tu_0^t
    -\left(2au_0^tu_0^{\phi}-a^2u_0^{\phi}u_0^{\phi}\right)
    -r^3M^{-1}u_0^{\phi}u_0^{\phi}\right]
    \label{eq:u1-case2}
\end{equation}
In the equation above, the 1st term on the RHS corresponds to the drift induced by gravity (or BH mass) in the Newtonian scheme. The 2nd term corresponds to the drift induced by the gravitomagnetic field, namely the drift induced by BH rotation, in the Newtonian scheme. It consists of terms linear and quadratic in $a u_0^{\phi}$. The linear term reduces to the 2nd term on the RHS of Eq.~\eqref{eq:vd-Newtonian3} in the limit $r\gg M$ and $u_0^t\rightarrow 1$ (hence $u_0^{\phi}\rightarrow 0$), while the quadratic term becomes effective only in the ultra-relativistic regime. The 3rd term corresponds to the drift induced by the curvature of magnetic field lines. It is identical to Eq.~\eqref{eq:vd-curvature} in the limit $r\gg M$ and $u_0^t\rightarrow 1$, for which the curvature of the toroidal magnetic field approaches $-\Vec{r}/r^2$. We emphasize that this term arises from the 2nd term, not the 1st term, on the RHS of Eq.~\eqref{eq:u1-2}, unlike what we discussed in Sect.~\ref{sec:limit}. This is because the $\phi$ axis in BL coordinates is intrinsically curved.

\section{Drift Rate}
\label{sec:rate}

The significance of drift is worth considering. Formally, we would like to study the ratio of the 1st order term to the leading term, $R_{\rm d}:=\left|\Vec{v}_{\rm d}\right|/\left|\Vec{v}_0\right|$. The drift is considered prominent when $R_{\rm d}$ takes a relatively substantial value ($\sim 0.1$), whereas the guiding-center approximation becomes invalid if $R_{\rm d}$ is too large ($\gtrsim 1$).

In the GR scheme, before assessing the effect of any physical phenomenon, one must specify the observer. For any observer, an orthonormal tetrad can be constructed. We set the 4-velocity of the observer as $e_{(t)}^{\mu}$ and the three spatial bases as $e_{(i)}^{\mu}$. For a moving particle with 4-velocity $u^{\mu}$, the observed Lorentz factor is $\hat{\gamma}=-u_{\mu} e_{(t)}^{\mu}$. The components of the observed 3-velocity are given by $\hat{v}_i=-u_{\mu} e_{(i)}^{\mu}/u_{\mu} e_{(t)}^{\mu}$. The validity of the guiding-center approximation requires $\hat{\gamma}\approx \hat{\gamma}_0= u^0_{\mu} e_{(t)}^{\mu}$. Thus we have:
\begin{equation}
    \hat{v}_{0,i} \approx -\frac{u^0_{\mu} e_{(i)}^{\mu}}{u^0_{\mu} e_{(t)}^{\mu}}~~,~~
    \hat{v}_{d,i} \approx -\frac{u^{1,\perp}_{\mu}~e_{(i)}^{\mu}}{u^0_{\mu} e_{(t)}^{\mu}}
    \label{eq:v0-v1}
\end{equation}
where $\hat{v}_{0,i}$ is the 0th order observed velocity of the charged particle, which is nearly along the magnetic field line, while $\hat{v}_{d,i}$ is the 1st order observed velocity, which corresponds to the drift (except for the electric induced one). It is natural to define the drift rate as:
\begin{equation}
    R_{\rm d} := \frac{\left|\hat{v}_{d,i}\right|}{\left|\hat{v}_{0,i}\right|} \equiv
    \frac{\left|u^{1,\perp}_{\mu}~e_{(i)}^{\mu}\right|}{\left|u^0_{\mu} e_{(i)}^{\mu}\right|}
    \label{eq:rate}
\end{equation}
This quantity represents the observed distance that a charged particle drifts while moving a unit length along the direction of magnetic field. Here the norm $\left|v_i\right|:=\sqrt{\sum_{i}{v_i^2}}$ is defined in the usual way.

Considering a drifting charged particle observed by the zero-angular-momentum observers (ZAMOs) in Kerr spacetime, whose tetrad takes the form:
\begin{equation}
    e_{(t)}^{\mu}=\frac{1}{\alpha}\left(\partial_t^{\mu}+\omega^{\phi}\partial_{\phi}^{\mu}\right)~~~,~~~
    e_{(i)}^{\mu}=\frac{1}{h_i}\partial_i^{\mu}
    \label{eq:ZAMOs}
\end{equation}
with:
\begin{equation}
    \begin{aligned}
        &\alpha=\sqrt{\frac{\Sigma\Delta}{A}}~~,~~
        \omega^{\phi}=\frac{2ar}{A}  \\
        &h_r=\sqrt{\frac{\Sigma}{\Delta}}~~,~~
        h_{\theta}=\sqrt{\Sigma}~~,~~
        h_{\phi}=\sqrt{\frac{A}{\Sigma}}\sin\theta
    \end{aligned}
\end{equation}
We keep $e$ positive in the following discussions. For \textit{case I}, where the magnetic field is purely radial on the equatorial plane, the drift rate becomes:
\begin{equation}
    R_{\rm d} = \frac{mM}{eB}\frac{2a\hat{\gamma}_0}{r^2}\sqrt{\frac{A~\hat{v}_{0,r}^2}{\left(\Delta-a^2\right)\left(r^2 \Delta~\hat{v}_{0,r}^2+4J^2\right)}}
    \label{eq:rate-case1}
\end{equation}
Here $\hat{\gamma}_0=1/\sqrt{1-\hat{v}_{0}^2}$ and $\hat{v}_{0,r}^2=\hat{v}_0^2-\frac{4J^2}{r^2\Delta}$, with $\hat{v}_0\equiv \left|\hat{v}_{0,i}\right|$ the leading order speed of the guiding-center as seen by ZAMOs. The condition:
\begin{equation}
    r^2\Delta-4J^2\hat{v}_0^{-2} \geq 0
    \label{eq:condition}
\end{equation}
ensures the timelike nature of $u_0^{\mu}$ (a purely radially moving particle does not exist everywhere in Kerr spacetime). For \textit{case II}, where the magnetic field is purely toroidal on the equatorial plane, we have:
\begin{equation}
    R_{\rm d} = \frac{mM}{eB}\frac{A\hat{\gamma}_0}{r^5\sqrt{\Delta}~\left|\hat{v}_{0}\right|}\left|1-2av_0^{\phi}+a^2\left(v_0^{\phi}\right)^2-r^3M^{-1}\left(v_0^{\phi}\right)^2\right|
    \label{eq:rate-case2}
\end{equation}
with $v_0^{\phi}=\frac{r^2\sqrt{\Delta}}{A}\hat{v}_{0}+\frac{2Jr}{A}$. In this situation, three effects contribute to the drift. Accordingly, we define:
\begin{equation}
    \begin{aligned}
        R_{{\rm d},M} :=& \frac{mM}{eB}\frac{A\hat{\gamma}_0}{r^5\sqrt{\Delta}~\left|\hat{v}_{0}\right|} \\
        ~~~~~~~~~~~~~~\\
        R_{{\rm d},J} :=& \frac{mM}{eB}\frac{A\hat{\gamma}_0}{r^5\sqrt{\Delta}~\left|\hat{v}_{0}\right|}\left|-2av_0^{\phi}+a^2\left(v_0^{\phi}\right)^2\right| \\
        ~~~~~~~~~~~~~~\\
        R_{{\rm d},\Vec{\kappa}_B} :=& \frac{mM}{eB}\frac{M^{-1}A\hat{\gamma}_0\left(v_0^{\phi}\right)^2}{r^2\sqrt{\Delta}~\left|\hat{v}_{0}\right|}
    \end{aligned}
    \label{eq:rate-effects}
\end{equation}
as the gravity-induced, spin-induced and curvature-induced drift rates. Furthermore, the importance of spin-induced drift relative to the gravity-induced and curvature-induced ones could be assessed via the quantities:
\begin{equation}
    \hat{R}_{{\rm d},J-M} := \frac{R_{{\rm d},J}}{R_{{\rm d},J}+R_{{\rm d},M}} ~~~,~~~
    \hat{R}_{{\rm d},J-\Vec{\kappa}_B} := \frac{R_{{\rm d},J}}{R_{{\rm d},J}+R_{{\rm d},\Vec{\kappa}_B}}
\end{equation}
We may refer to $\hat{R}_{{\rm d},J-M}$ and $\hat{R}_{{\rm d},J-\Vec{\kappa}_B}$ as the relative spin-induced drift rates.

\begin{figure*}
    \centering
    \includegraphics[width=0.32\linewidth]{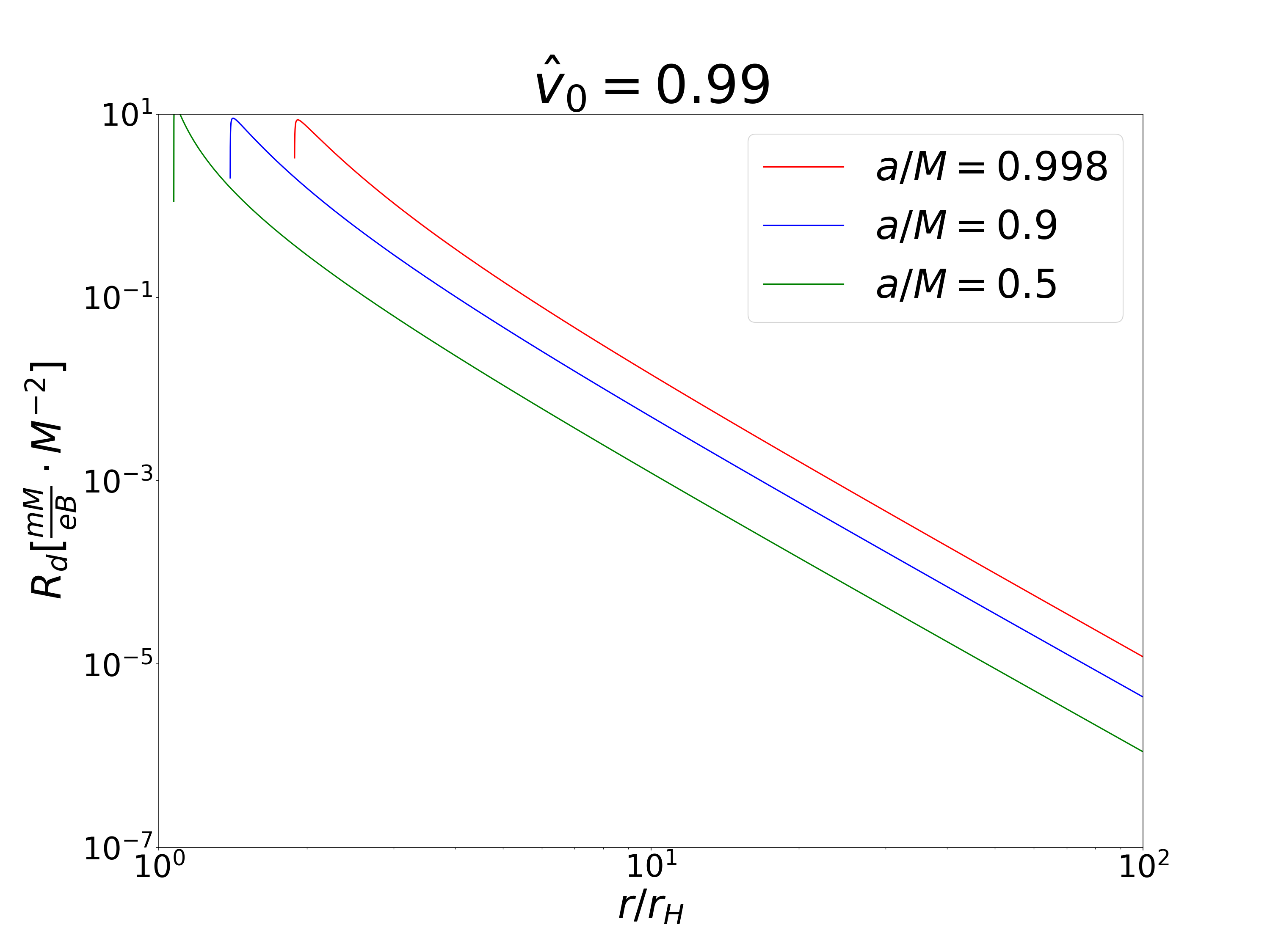}
    \includegraphics[width=0.32\linewidth]{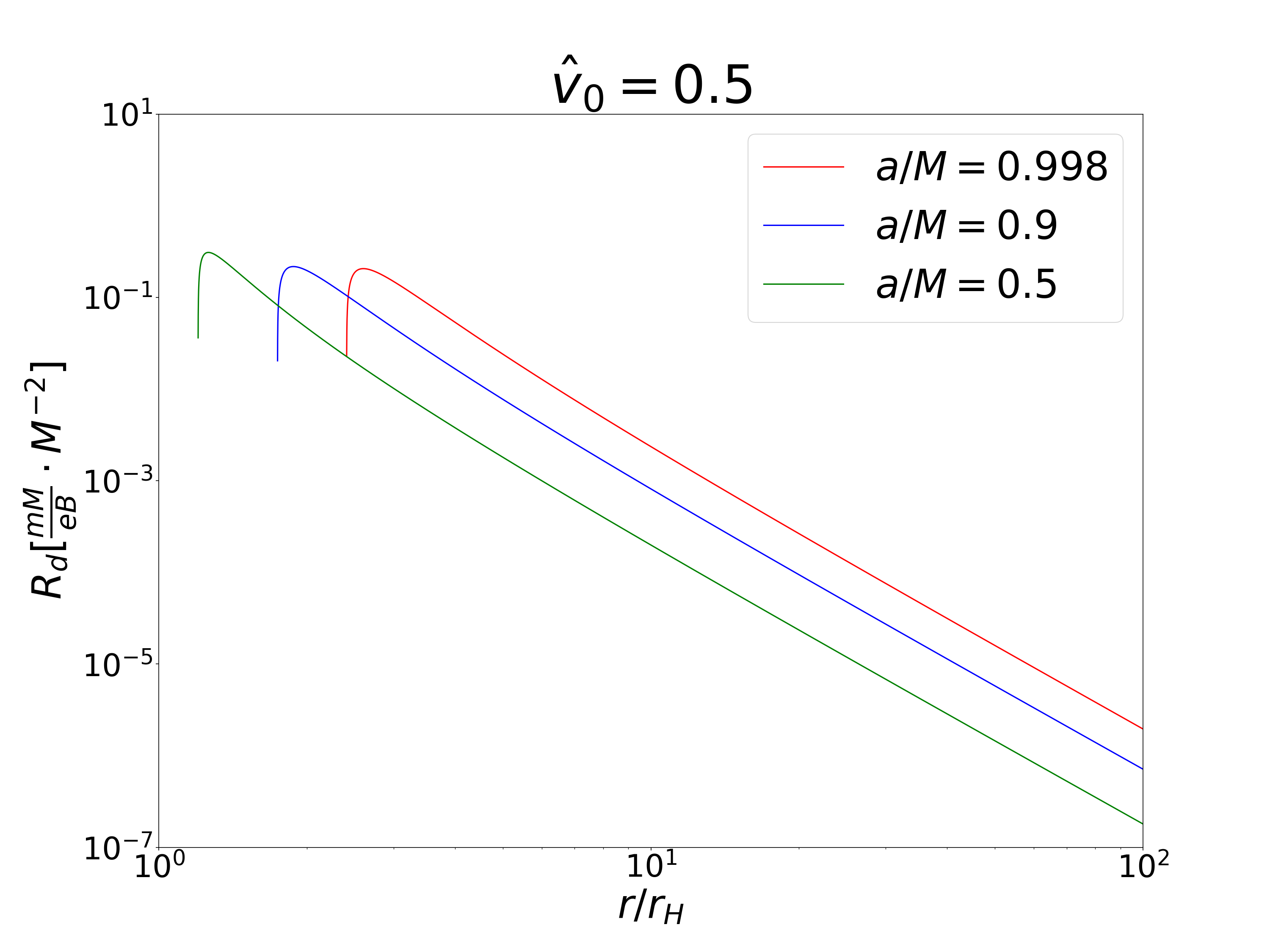}
    \includegraphics[width=0.32\linewidth]{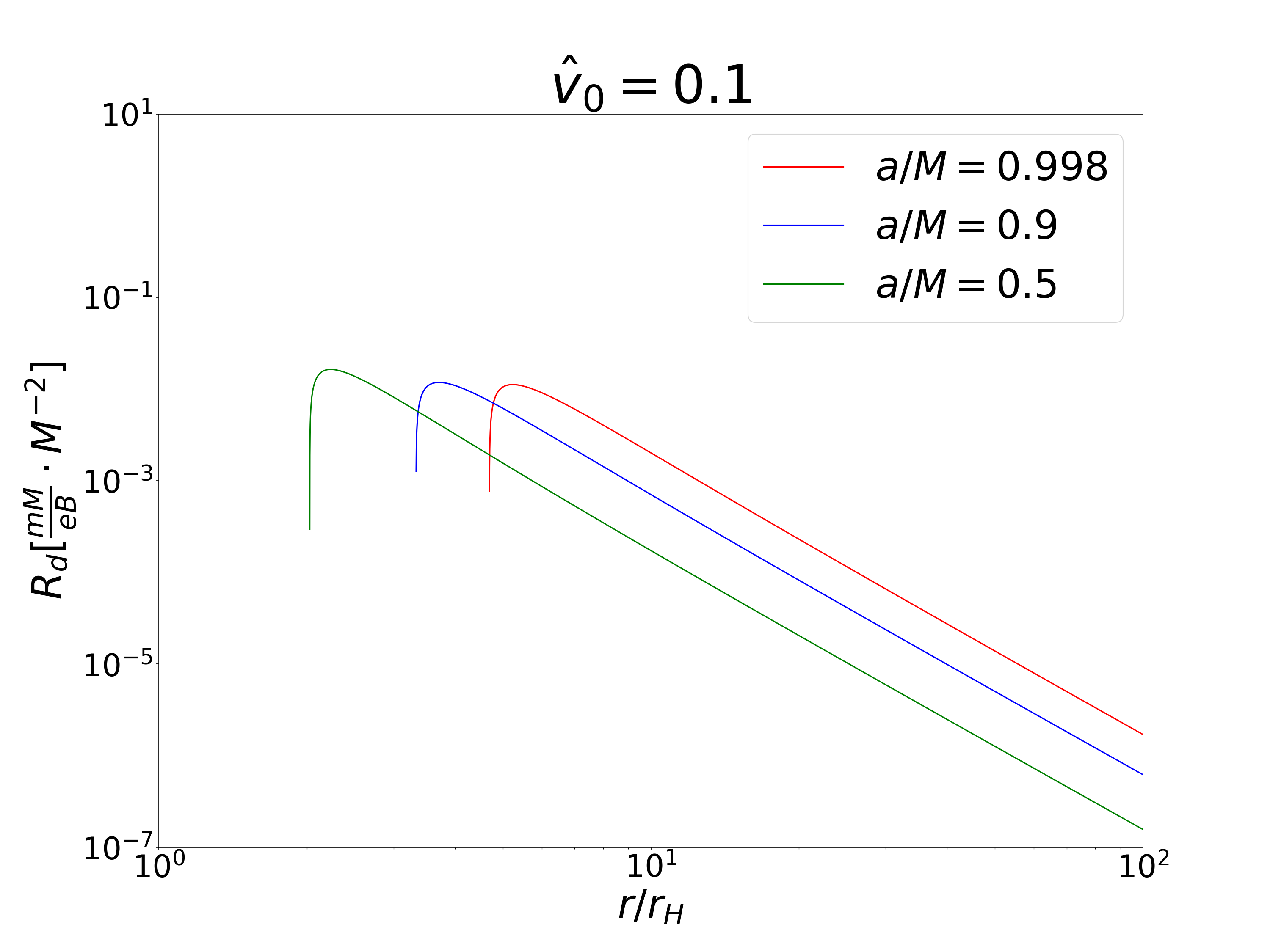}
    \includegraphics[width=0.32\linewidth]{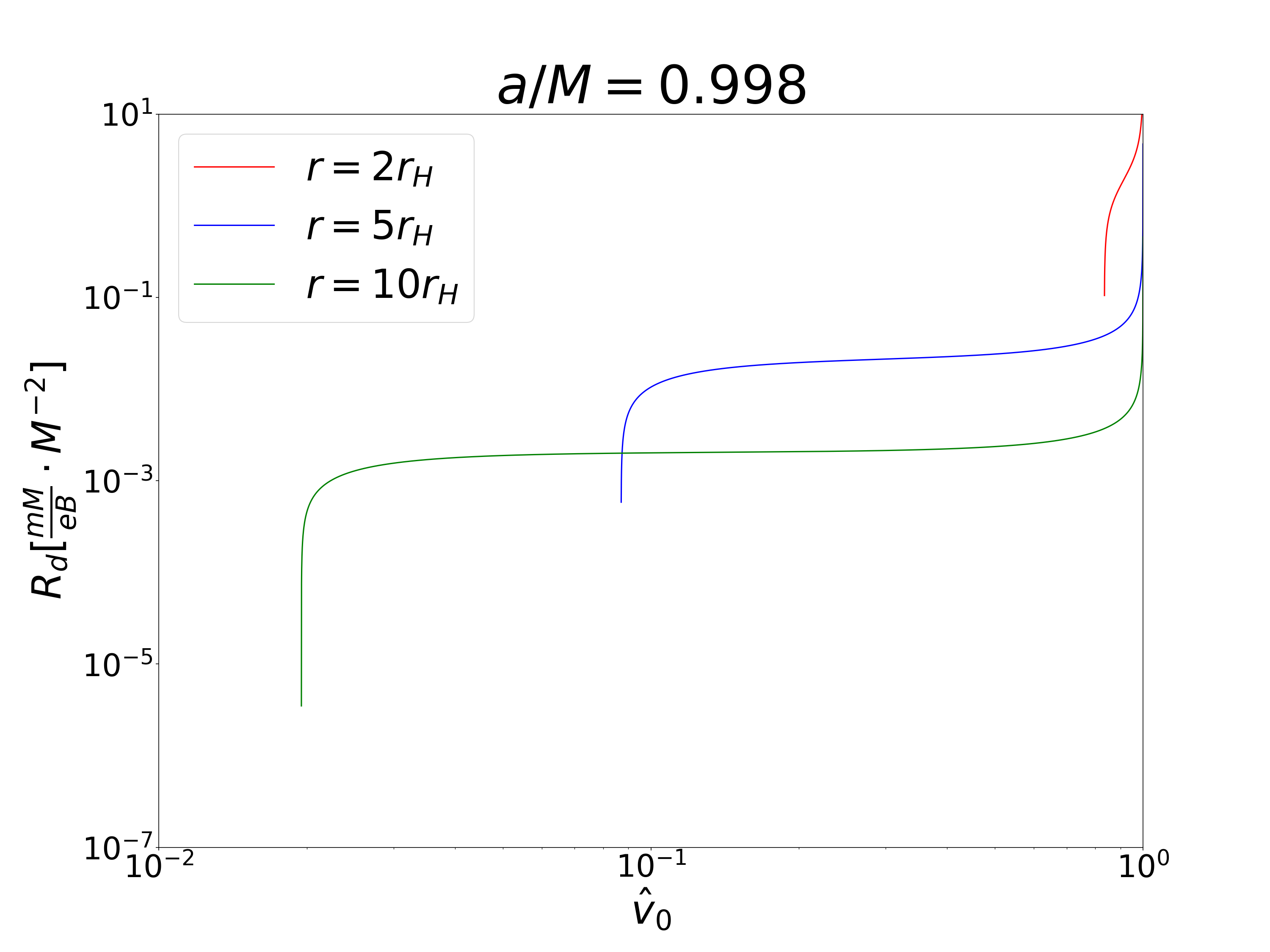}
    \includegraphics[width=0.32\linewidth]{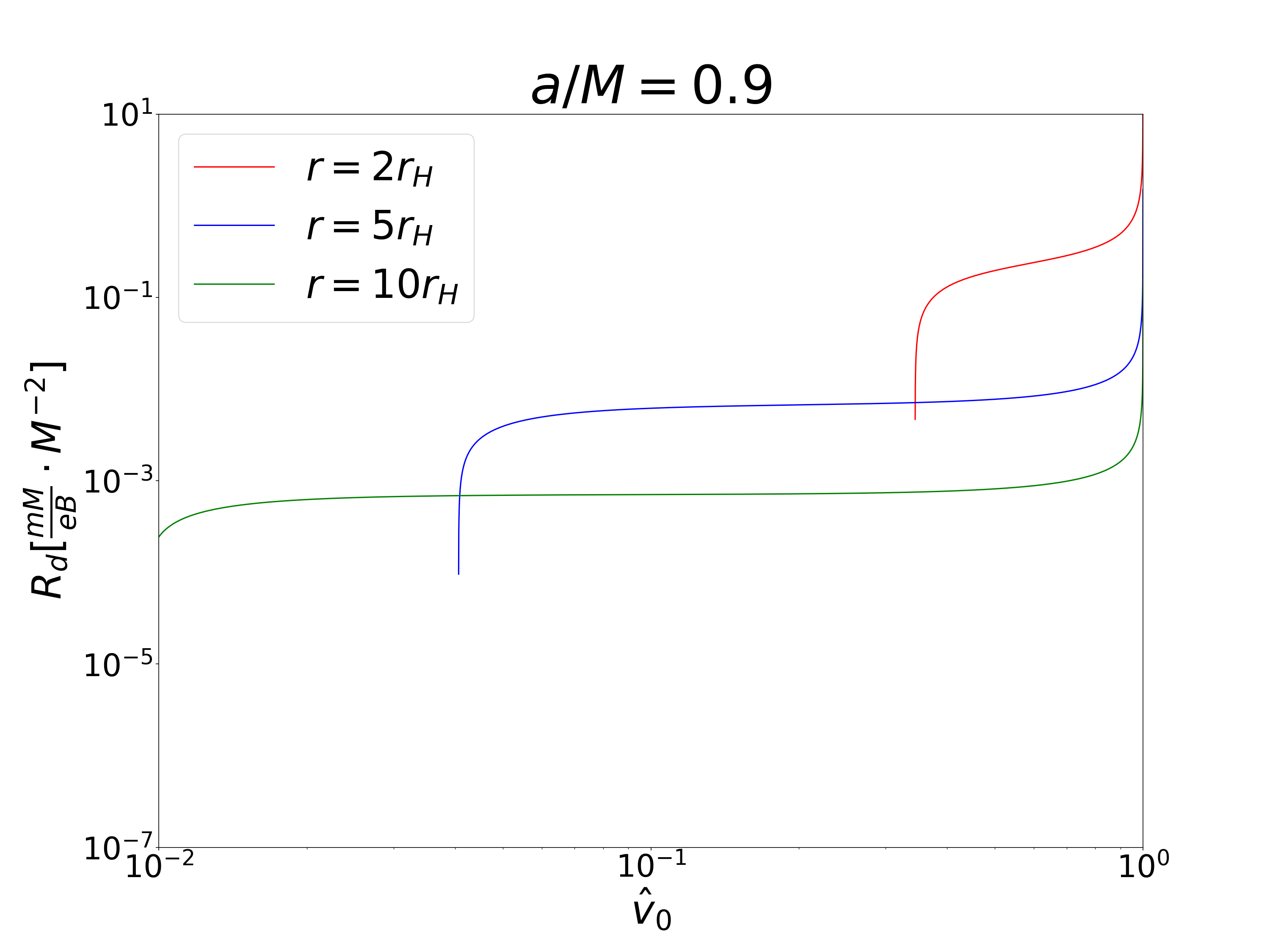}
    \includegraphics[width=0.32\linewidth]{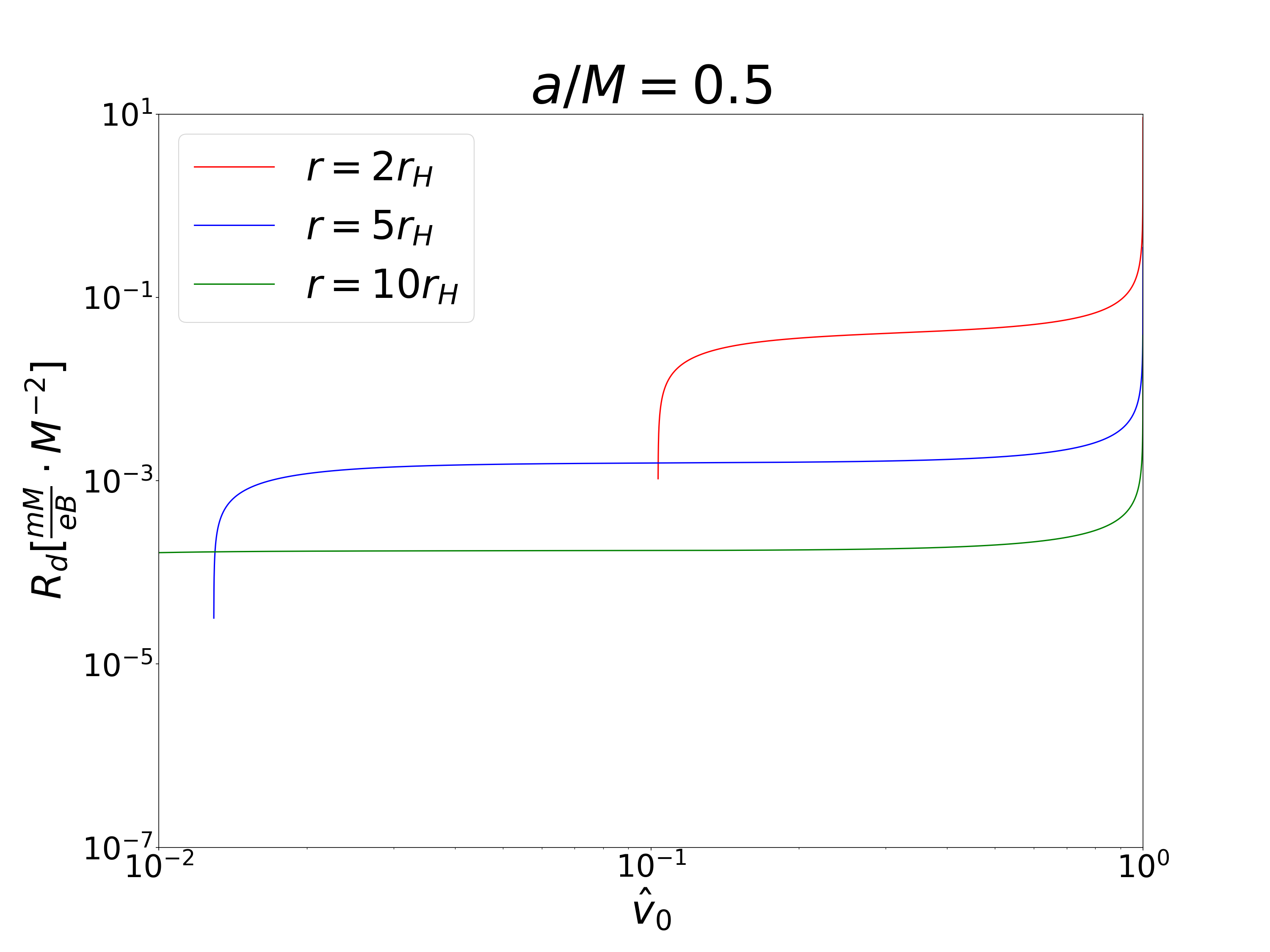}
    \includegraphics[width=0.32\linewidth]{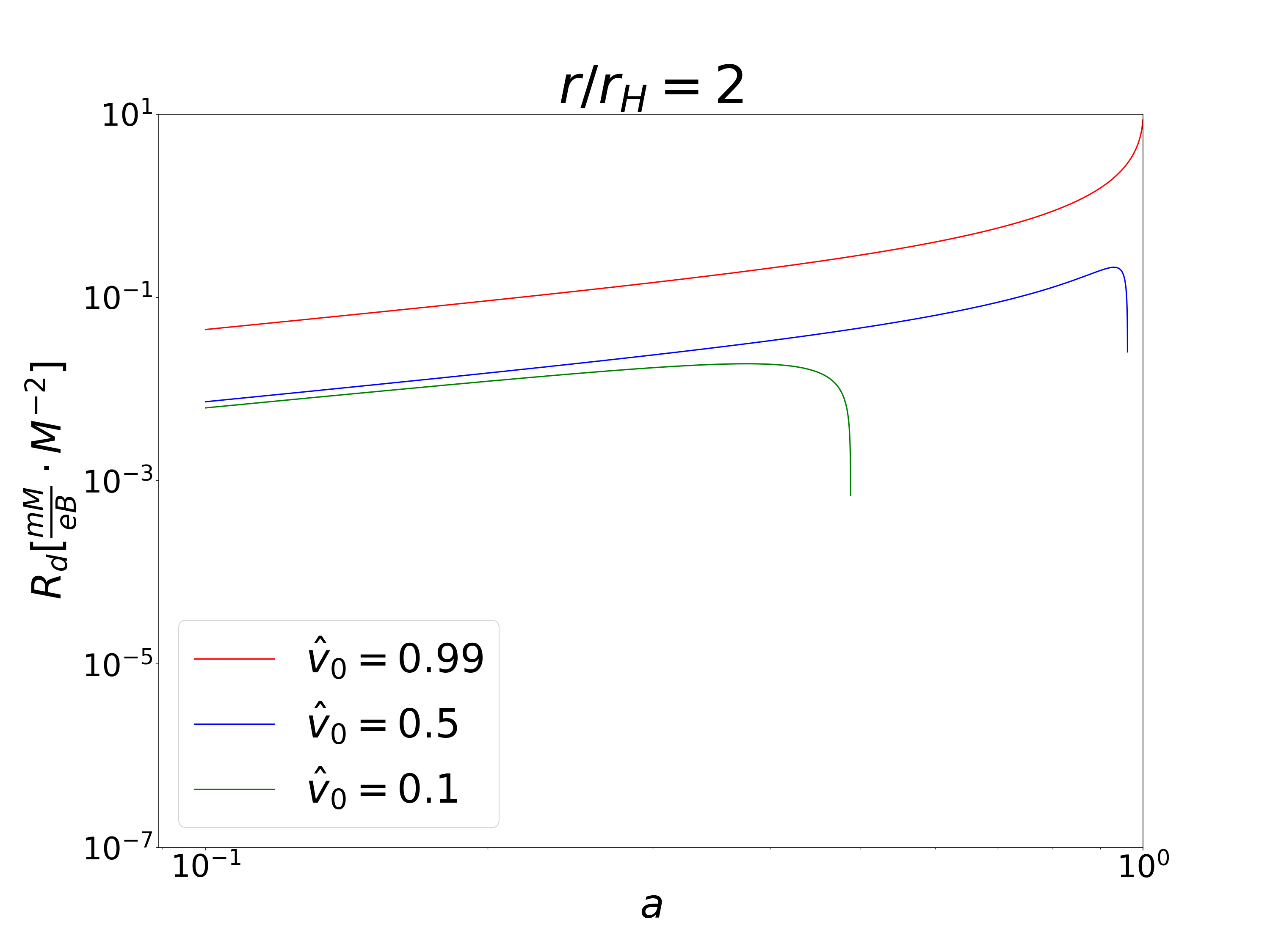}
    \includegraphics[width=0.32\linewidth]{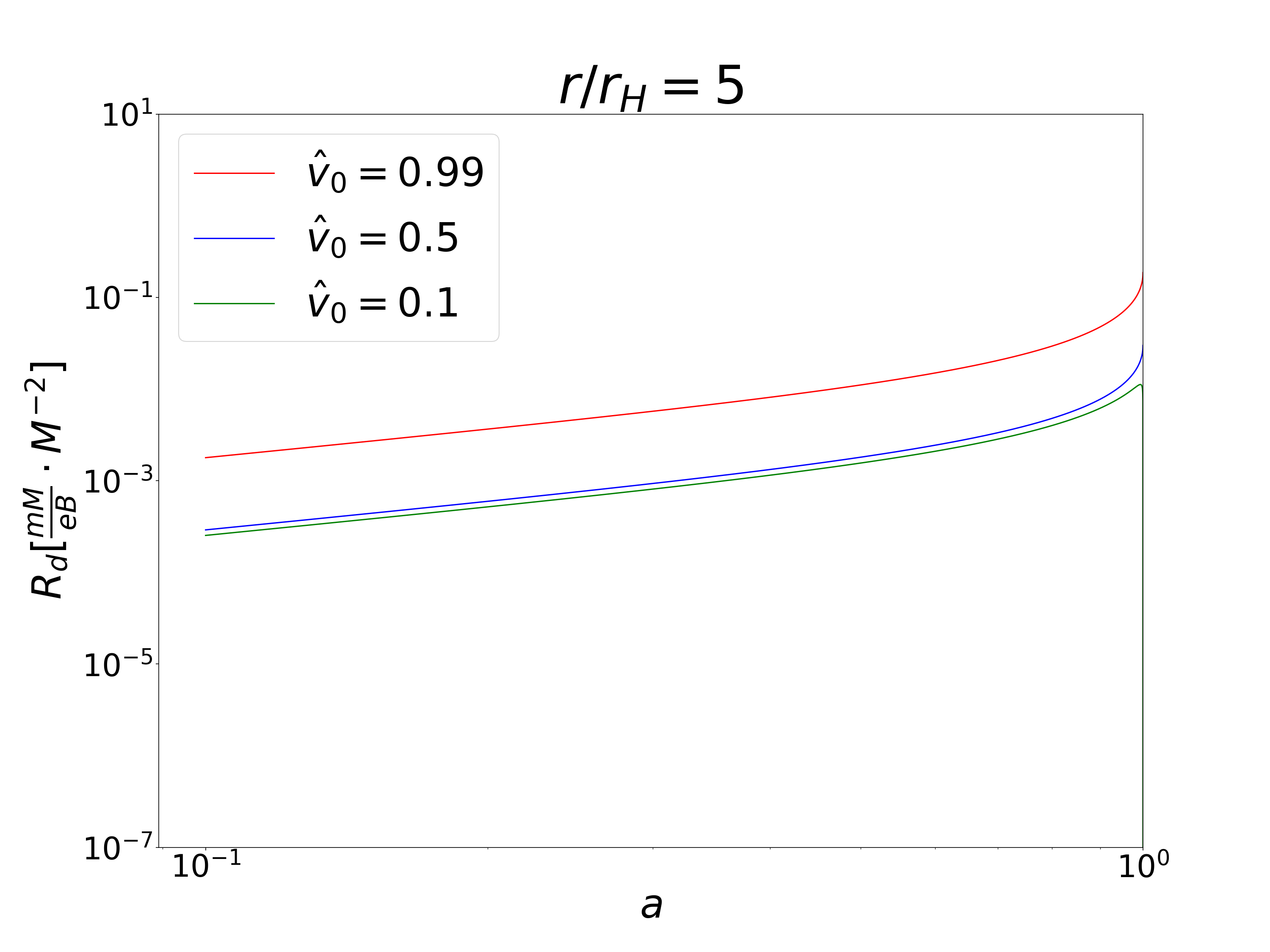}
    \includegraphics[width=0.32\linewidth]{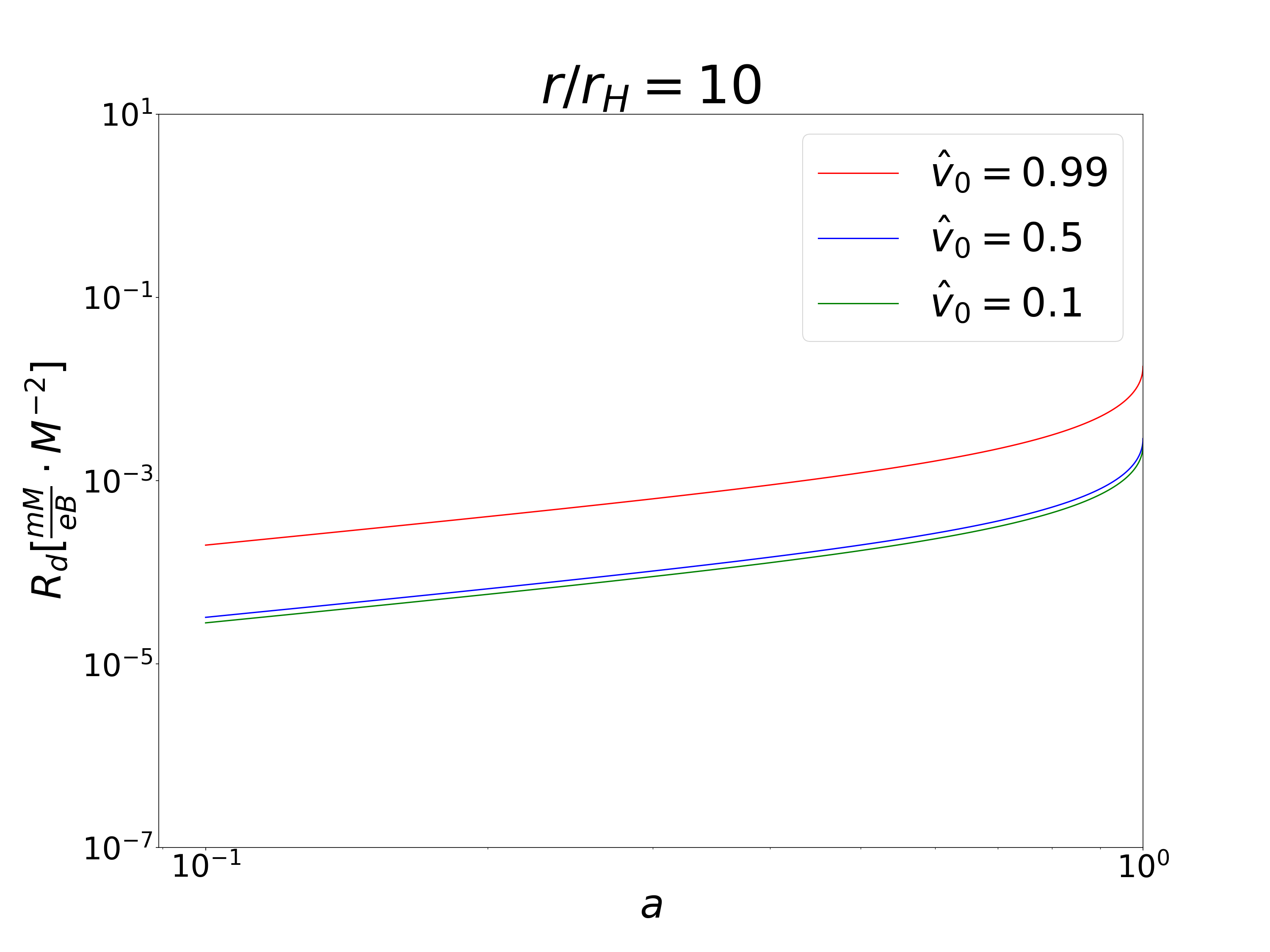}
    \caption{The variations of drift rate as functions of radius (top panels), 0th order observed velocity (middle panels) and BH spin (bottom panels) for \textit{case I}.}
    \label{fig:Rd1}
\end{figure*}

\begin{figure*}
    \centering
    \includegraphics[width=0.32\linewidth]{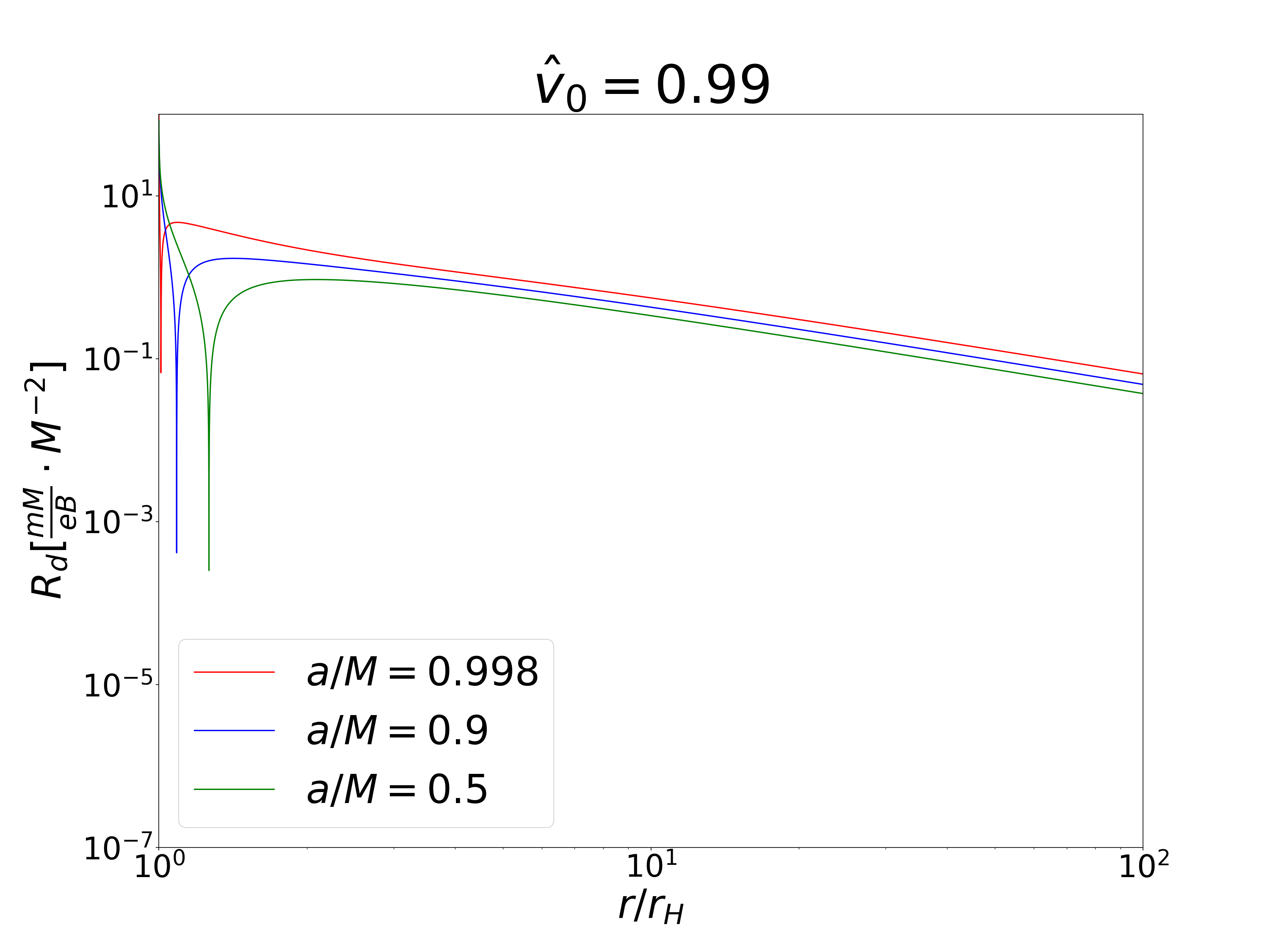}
    \includegraphics[width=0.32\linewidth]{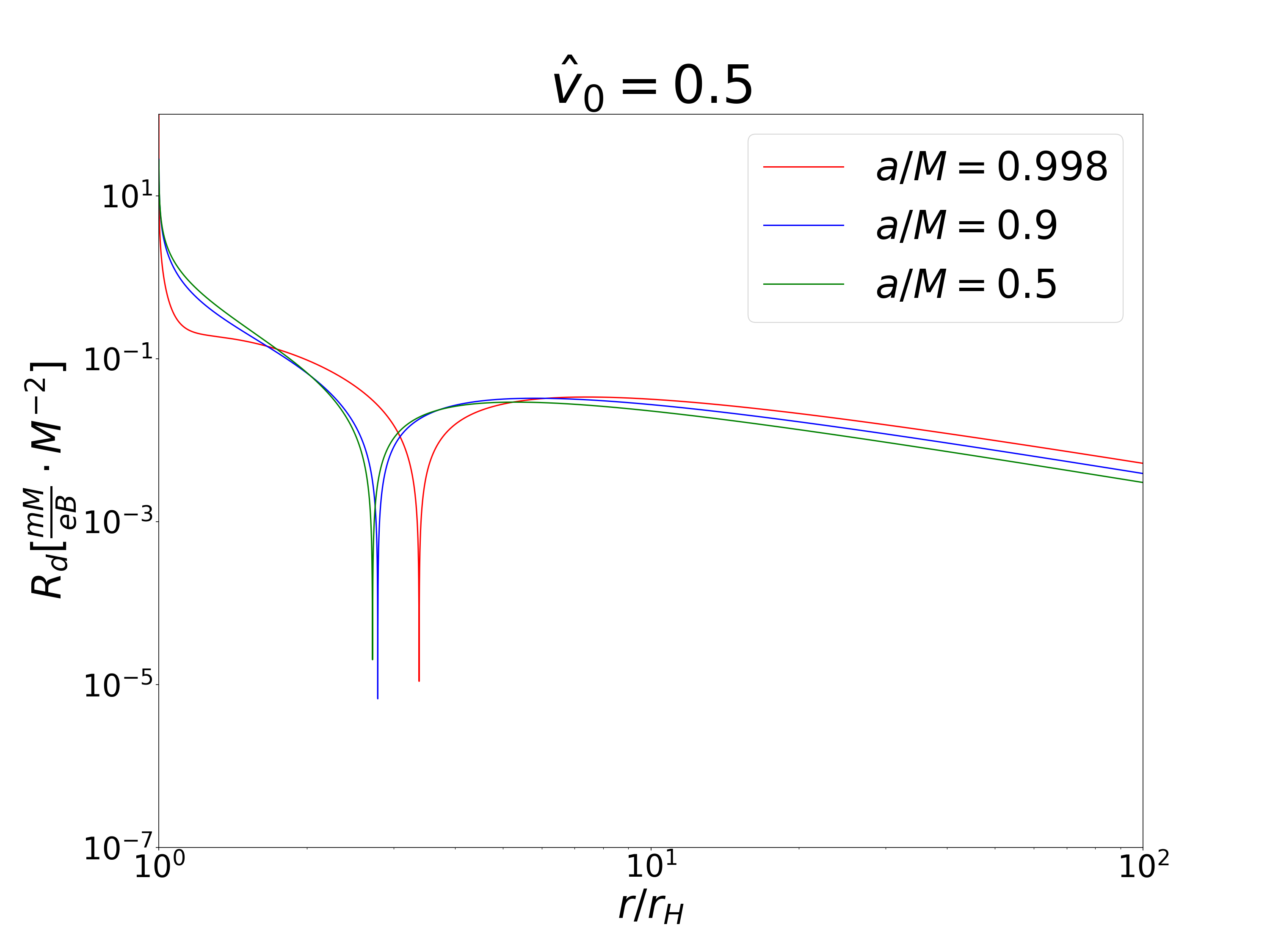}
    \includegraphics[width=0.32\linewidth]{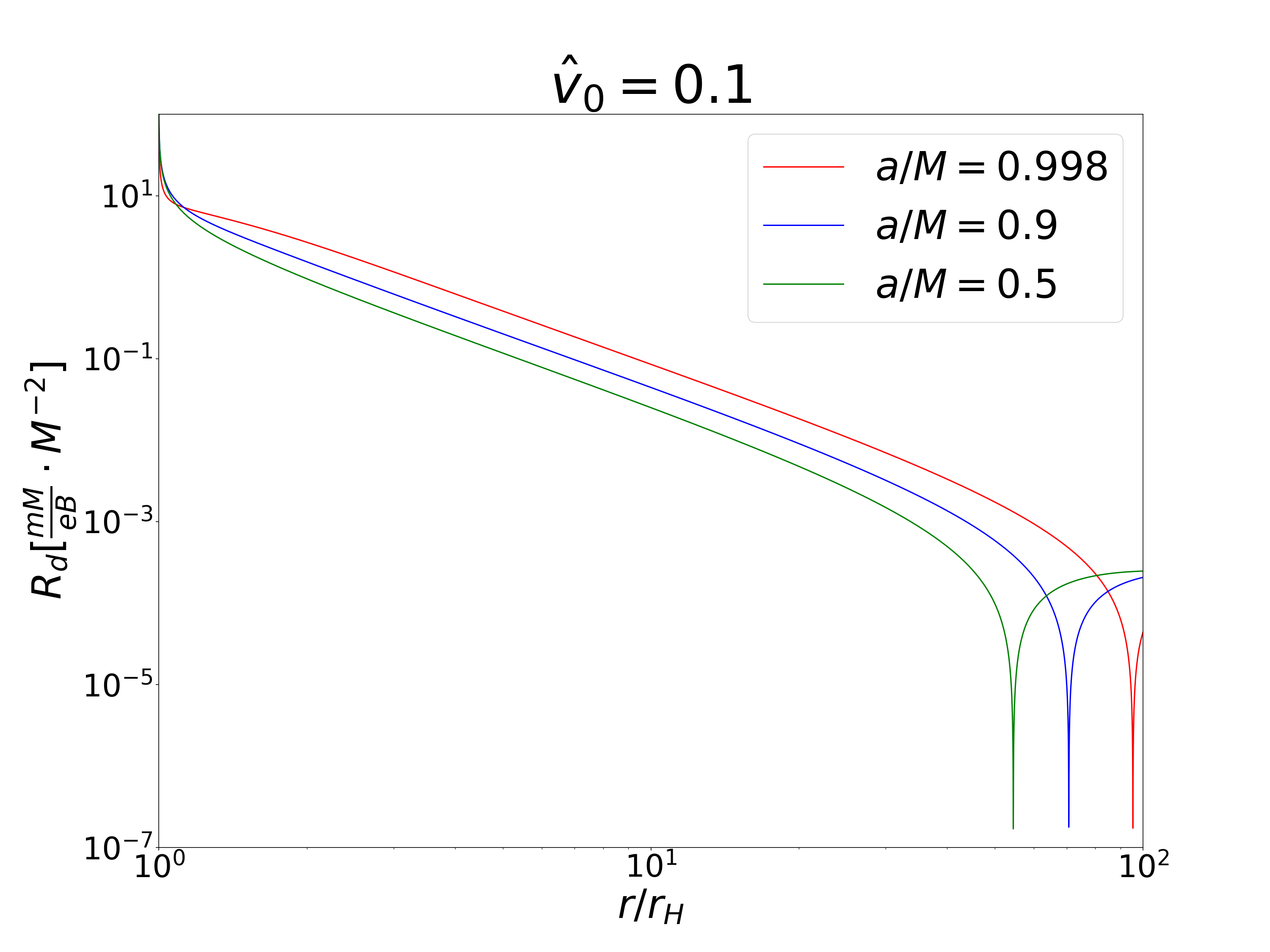}
    \includegraphics[width=0.32\linewidth]{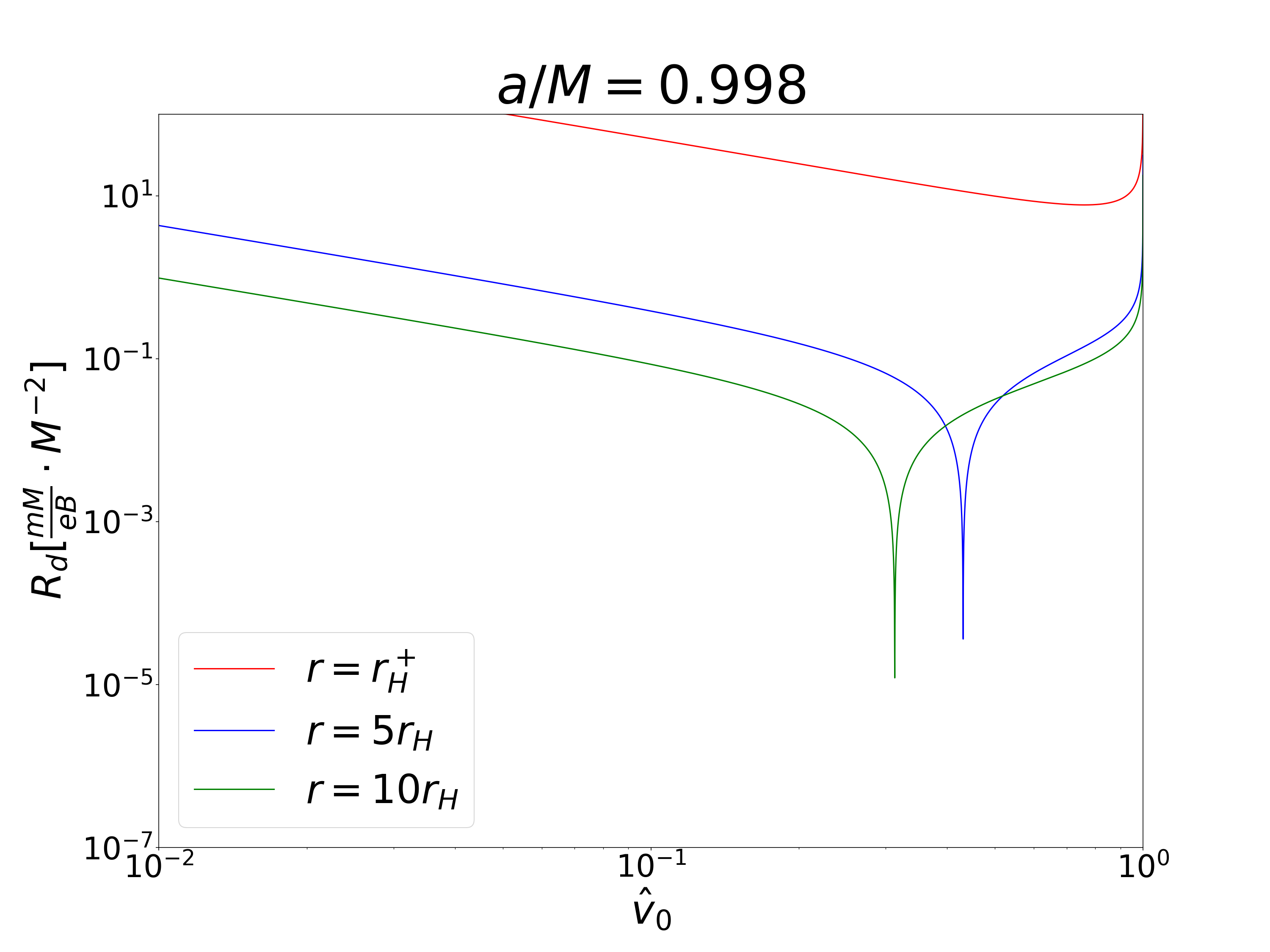}
    \includegraphics[width=0.32\linewidth]{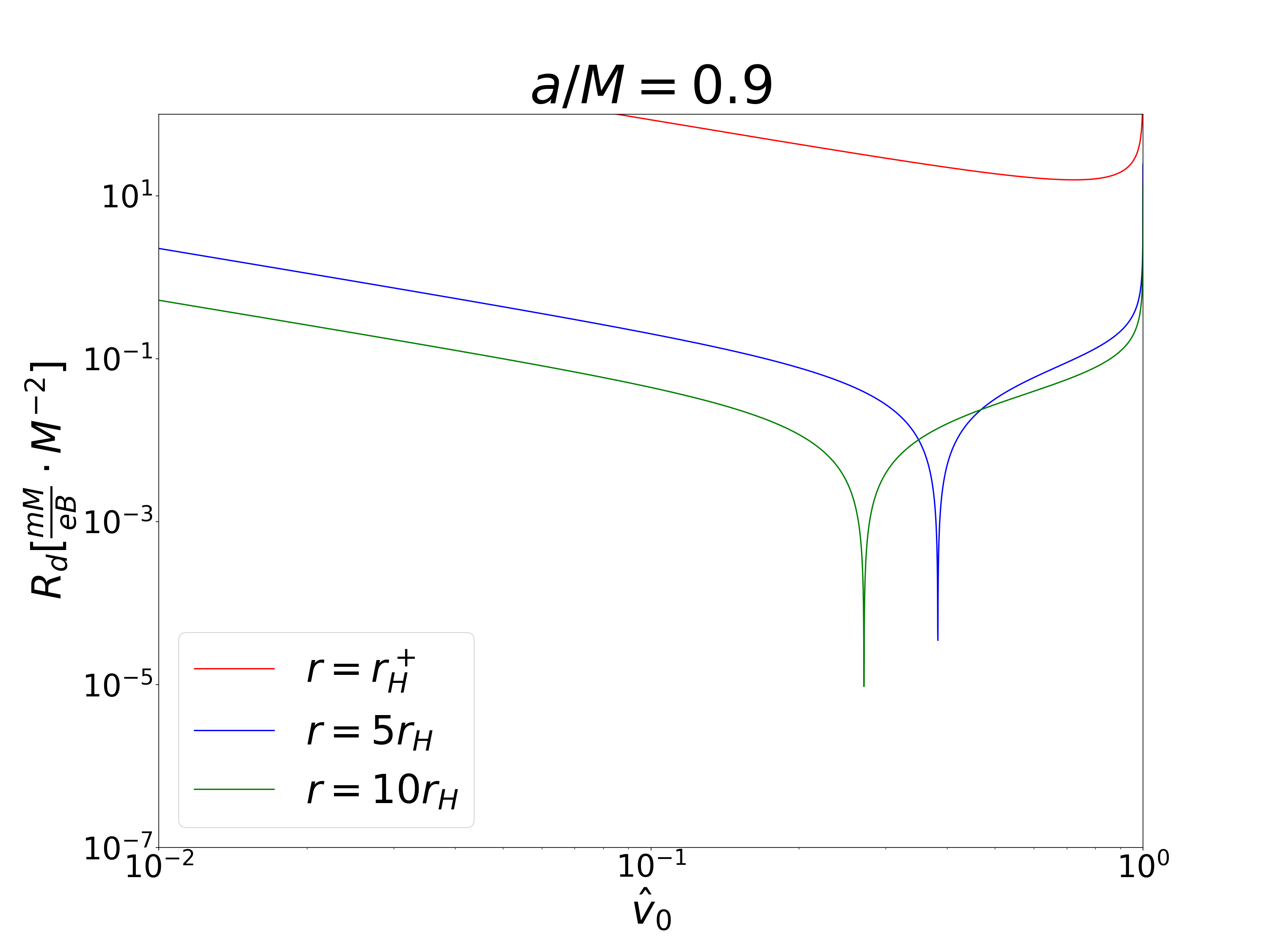}
    \includegraphics[width=0.32\linewidth]{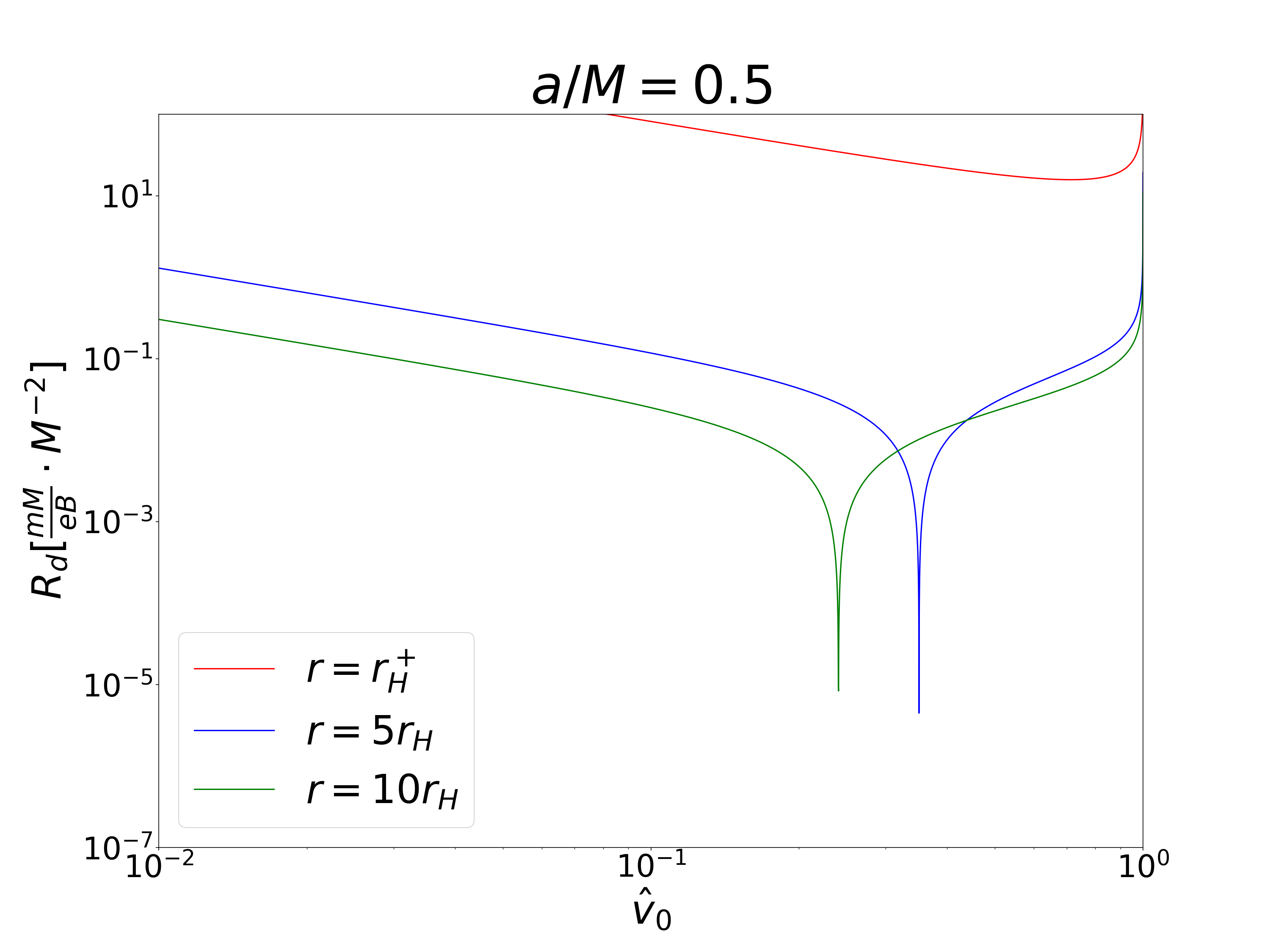}
    \includegraphics[width=0.32\linewidth]{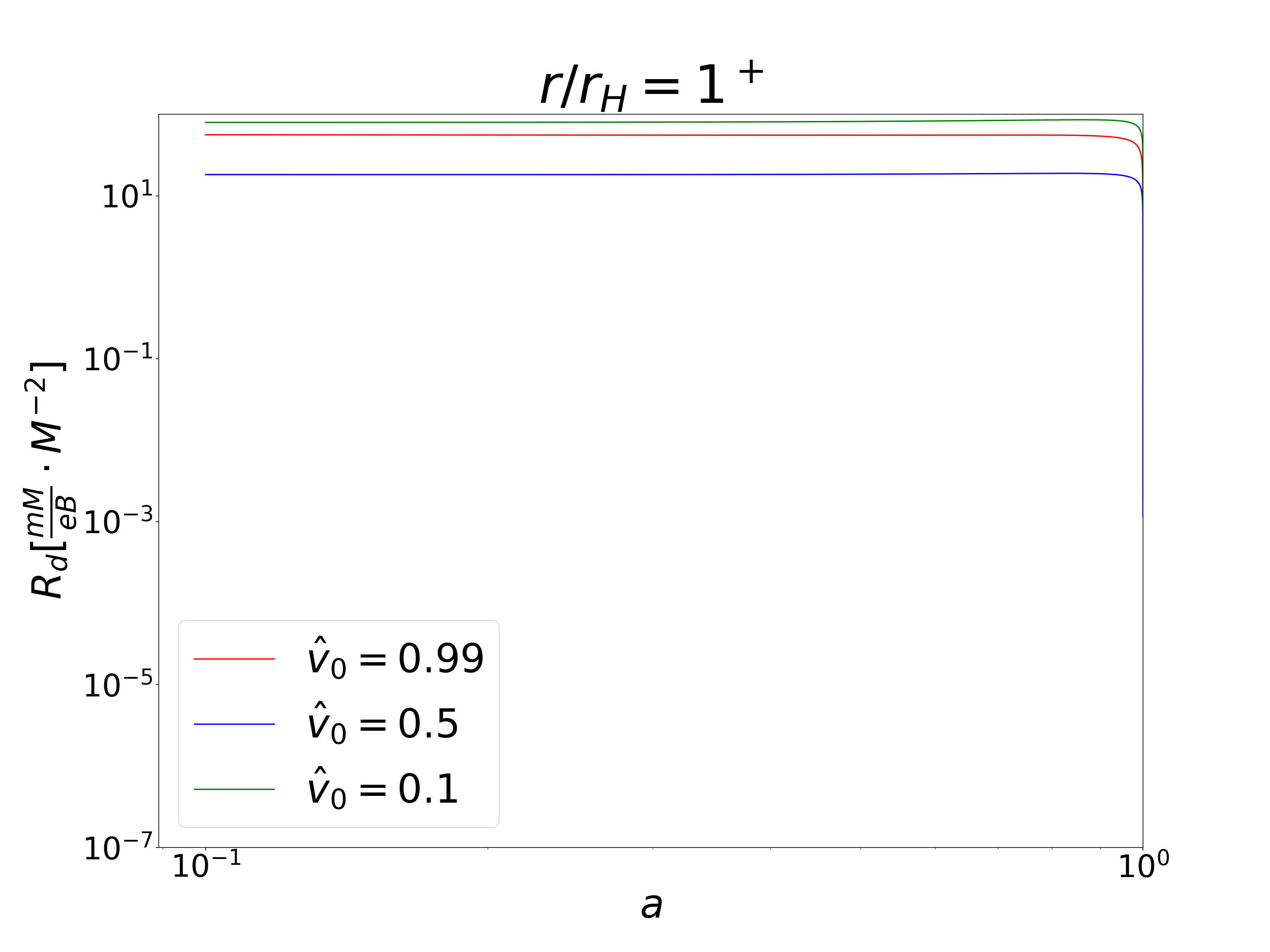}
    \includegraphics[width=0.32\linewidth]{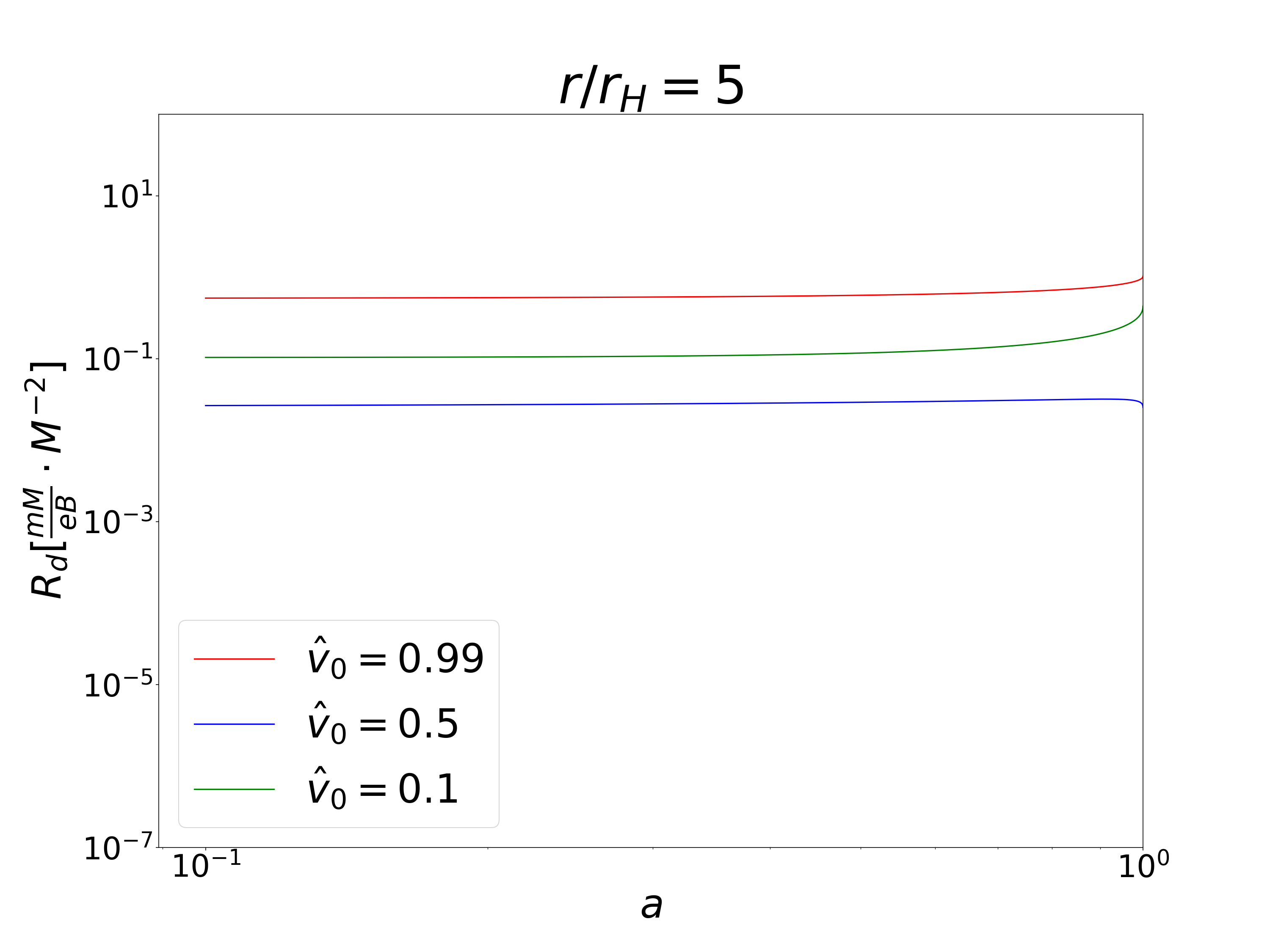}
    \includegraphics[width=0.32\linewidth]{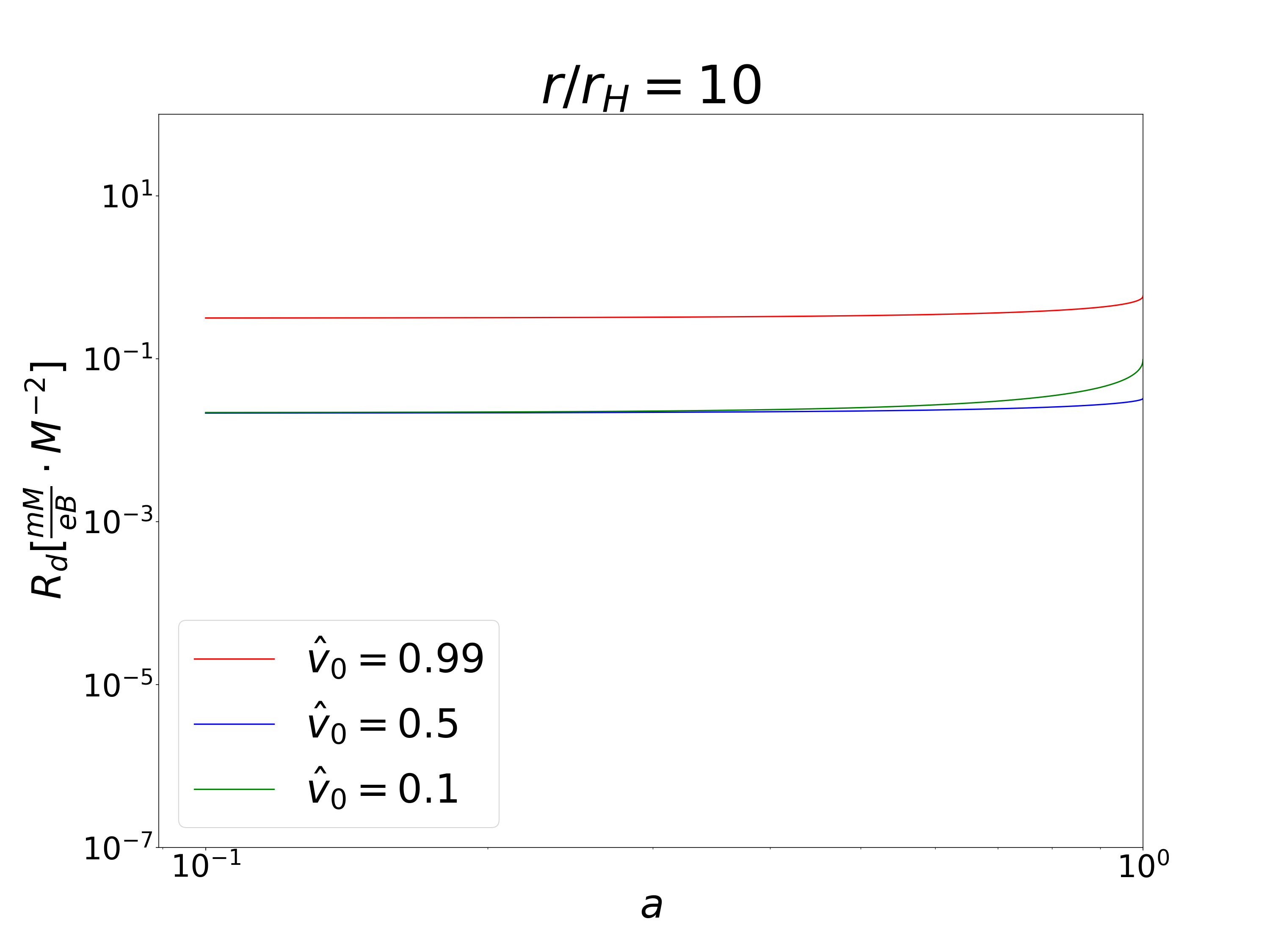}
    \caption{The variations of drift rate as functions of radius (top panels), 0th order observed velocity (middle panels) and BH spin (bottom panels) for \textit{case II}. Here $r/r_{\rm H}=1^+$ represents the radius outside but very close to the event horizon.}
    \label{fig:Rd2}
\end{figure*}

\begin{figure*}
    \centering
    \includegraphics[width=0.32\linewidth]{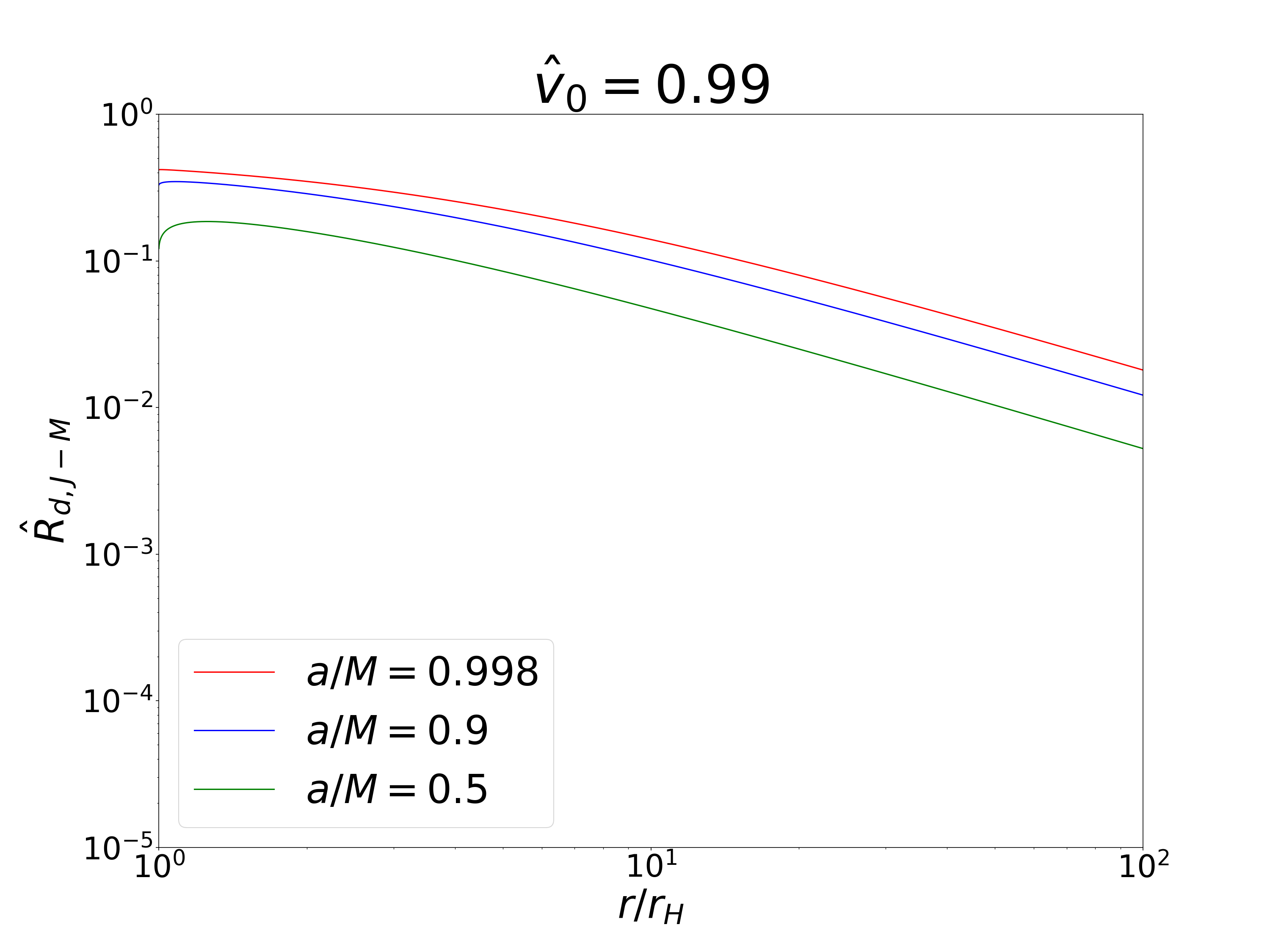}
    \includegraphics[width=0.32\linewidth]{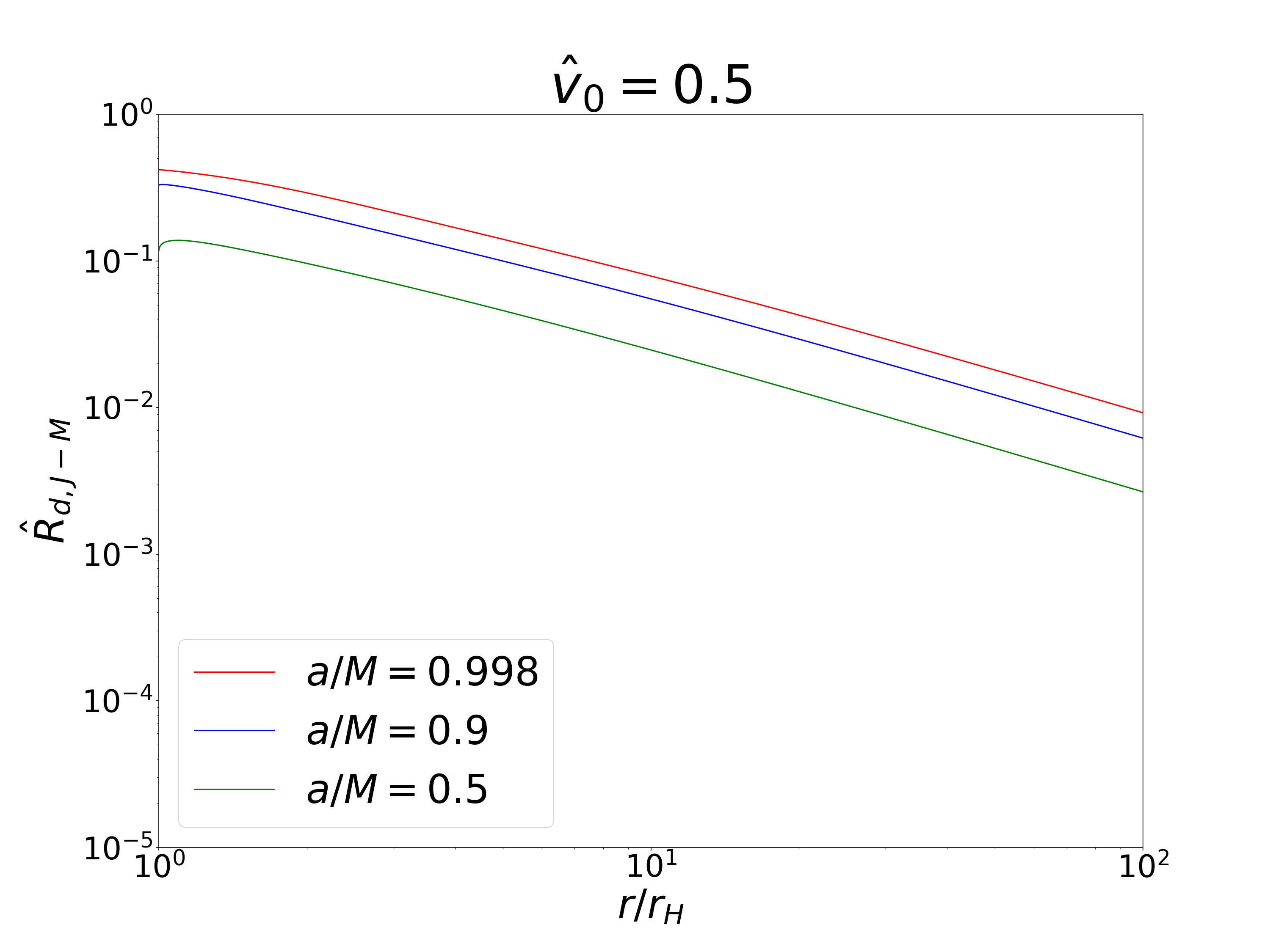}
    \includegraphics[width=0.32\linewidth]{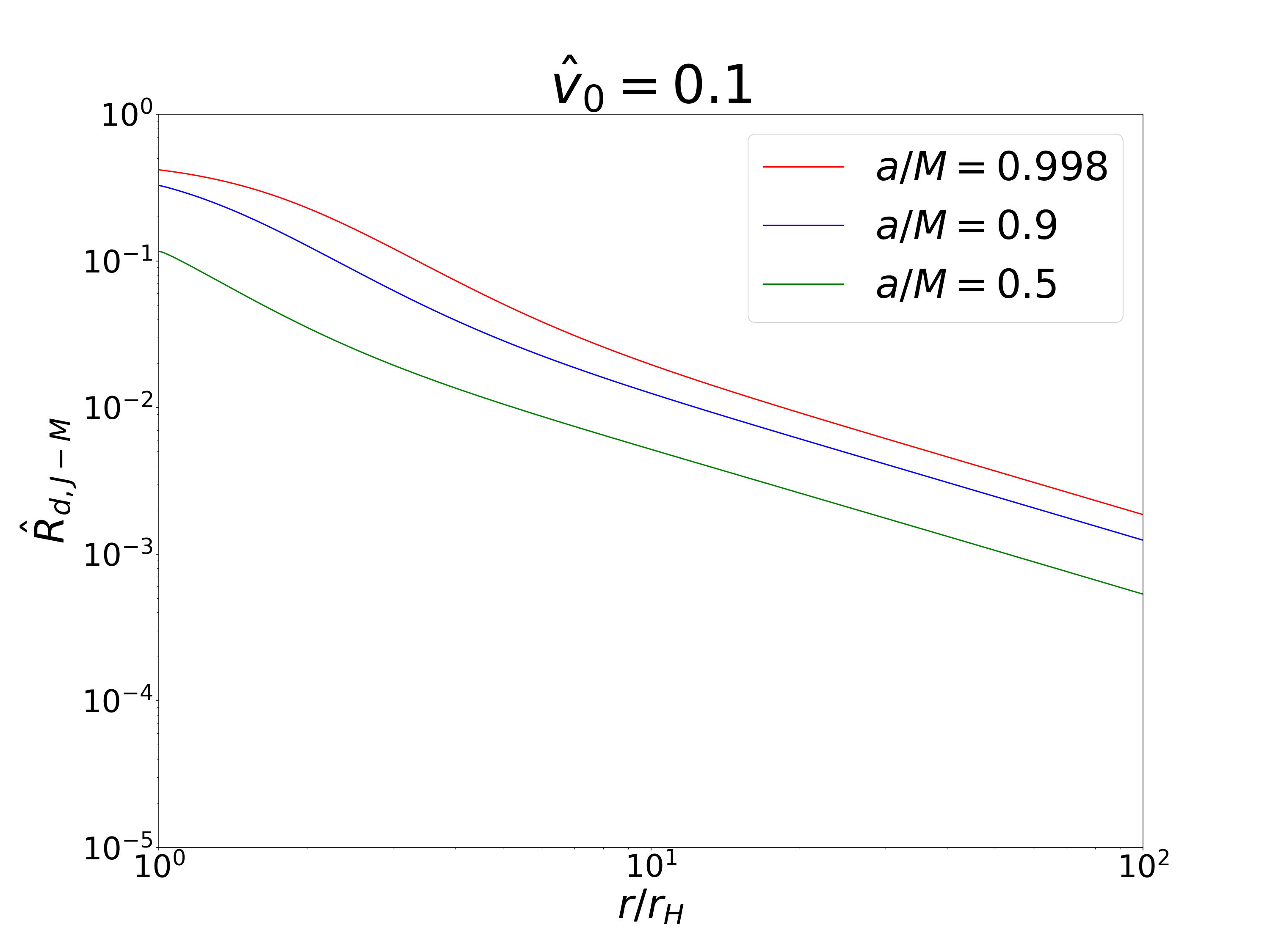}
    \includegraphics[width=0.32\linewidth]{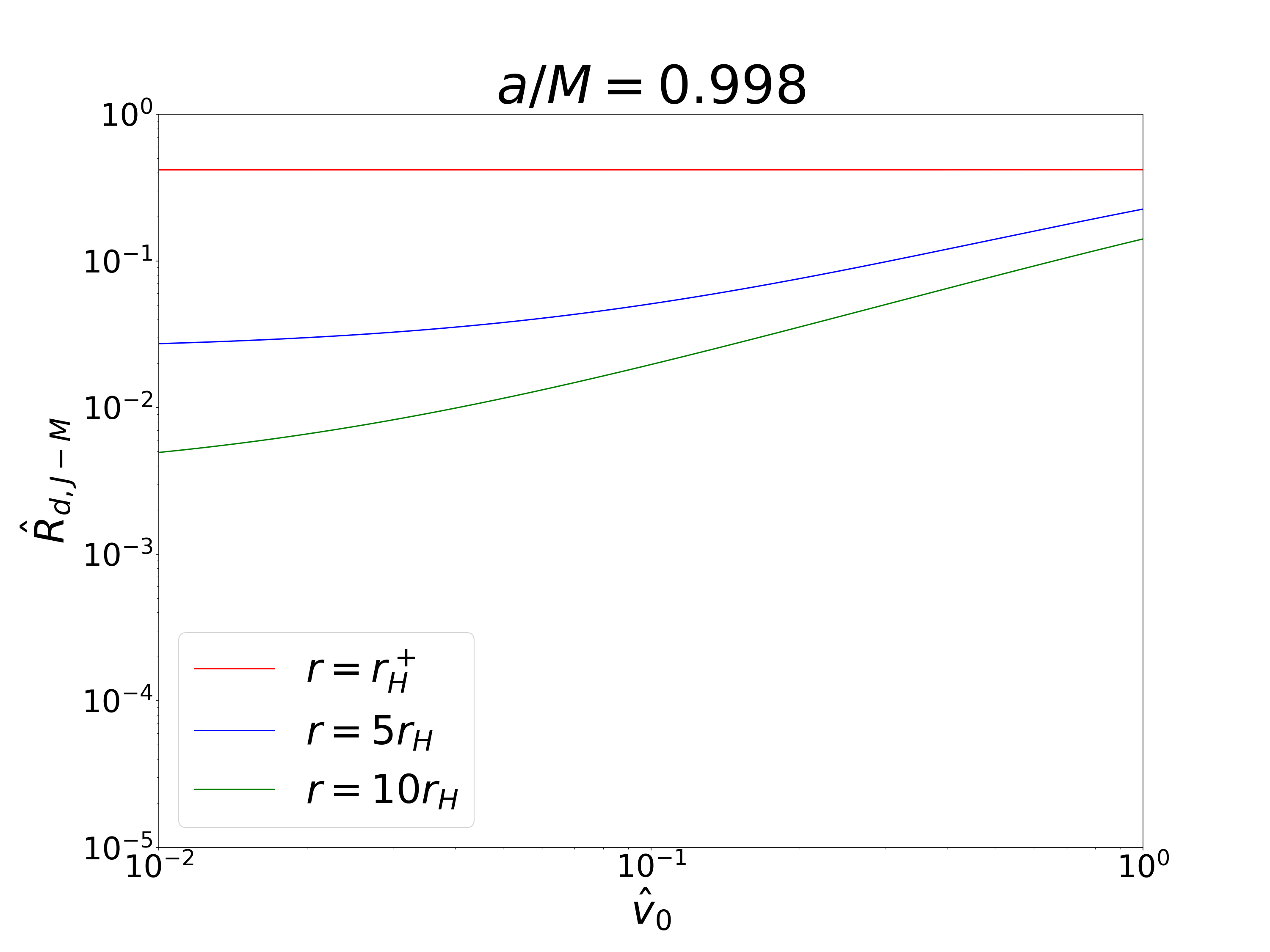}
    \includegraphics[width=0.32\linewidth]{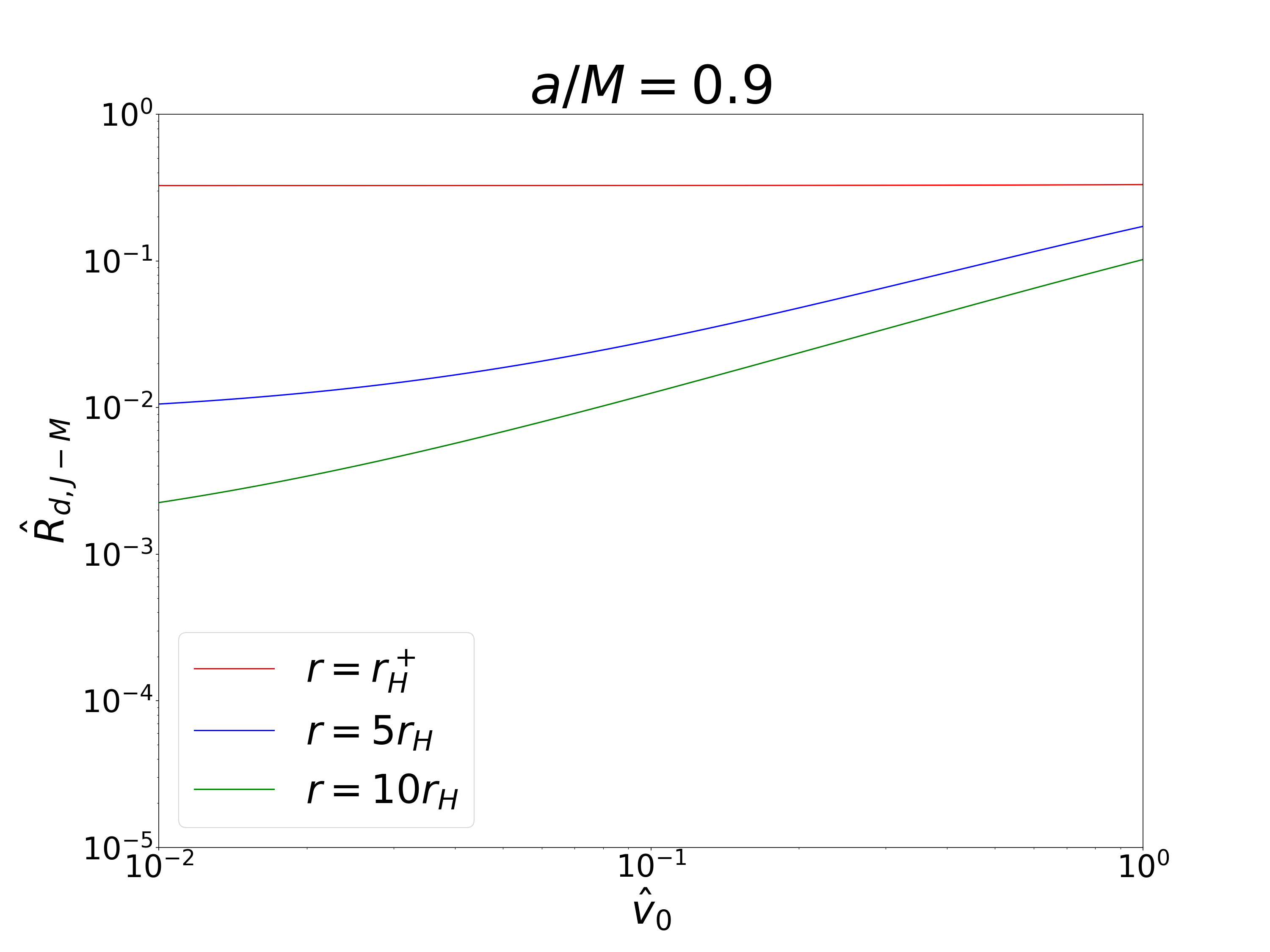}
    \includegraphics[width=0.32\linewidth]{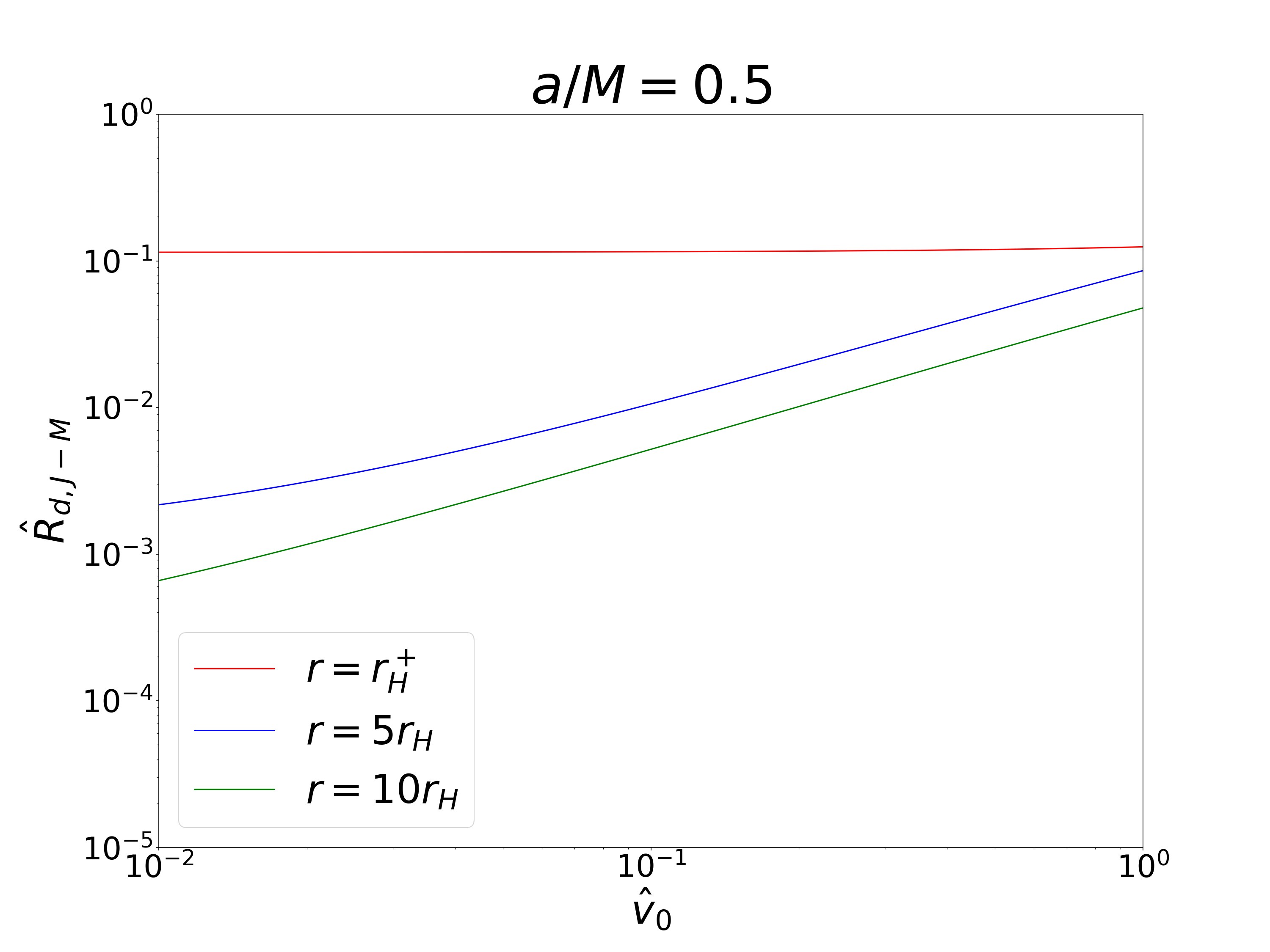}
    \includegraphics[width=0.32\linewidth]{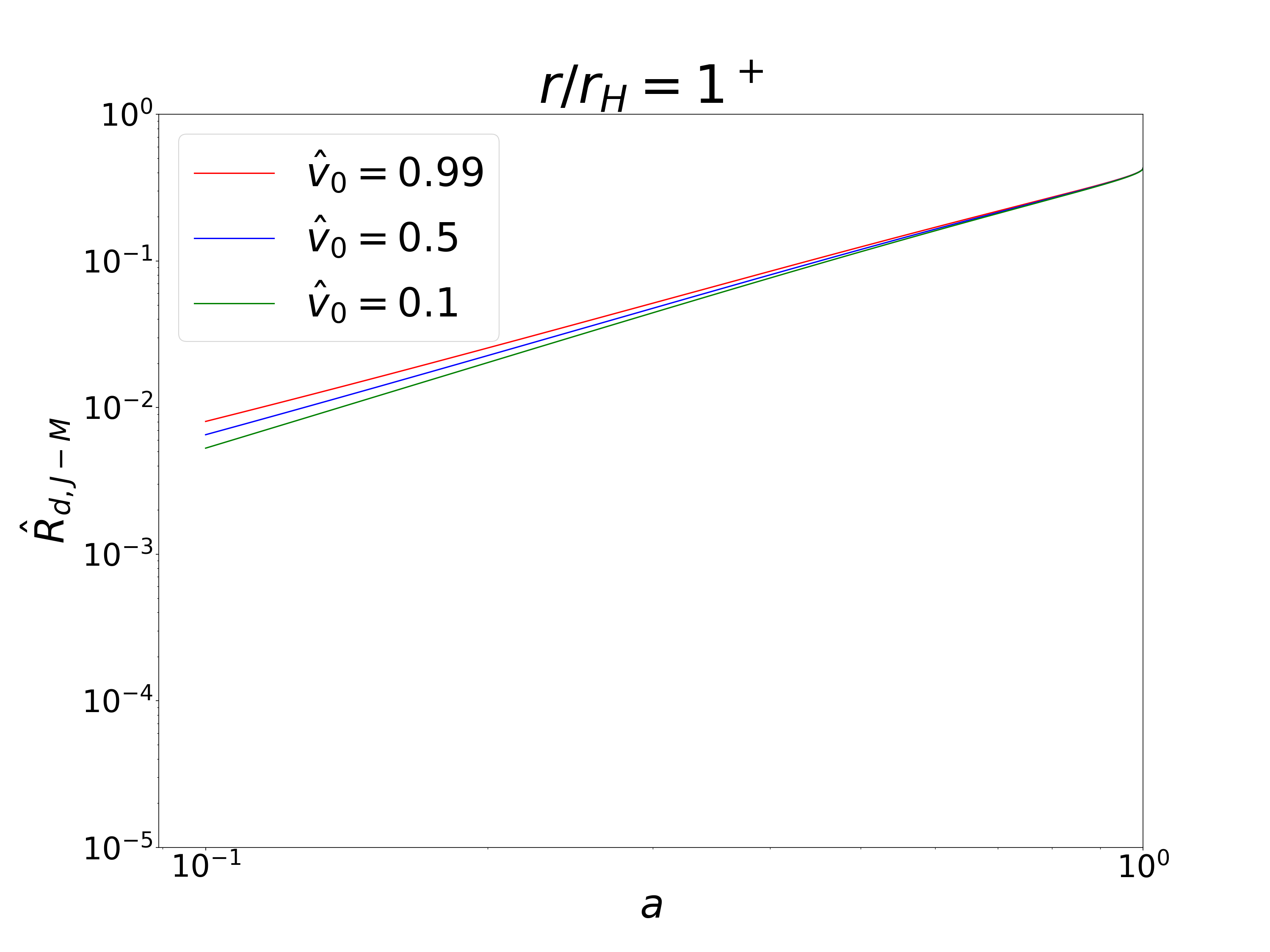}
    \includegraphics[width=0.32\linewidth]{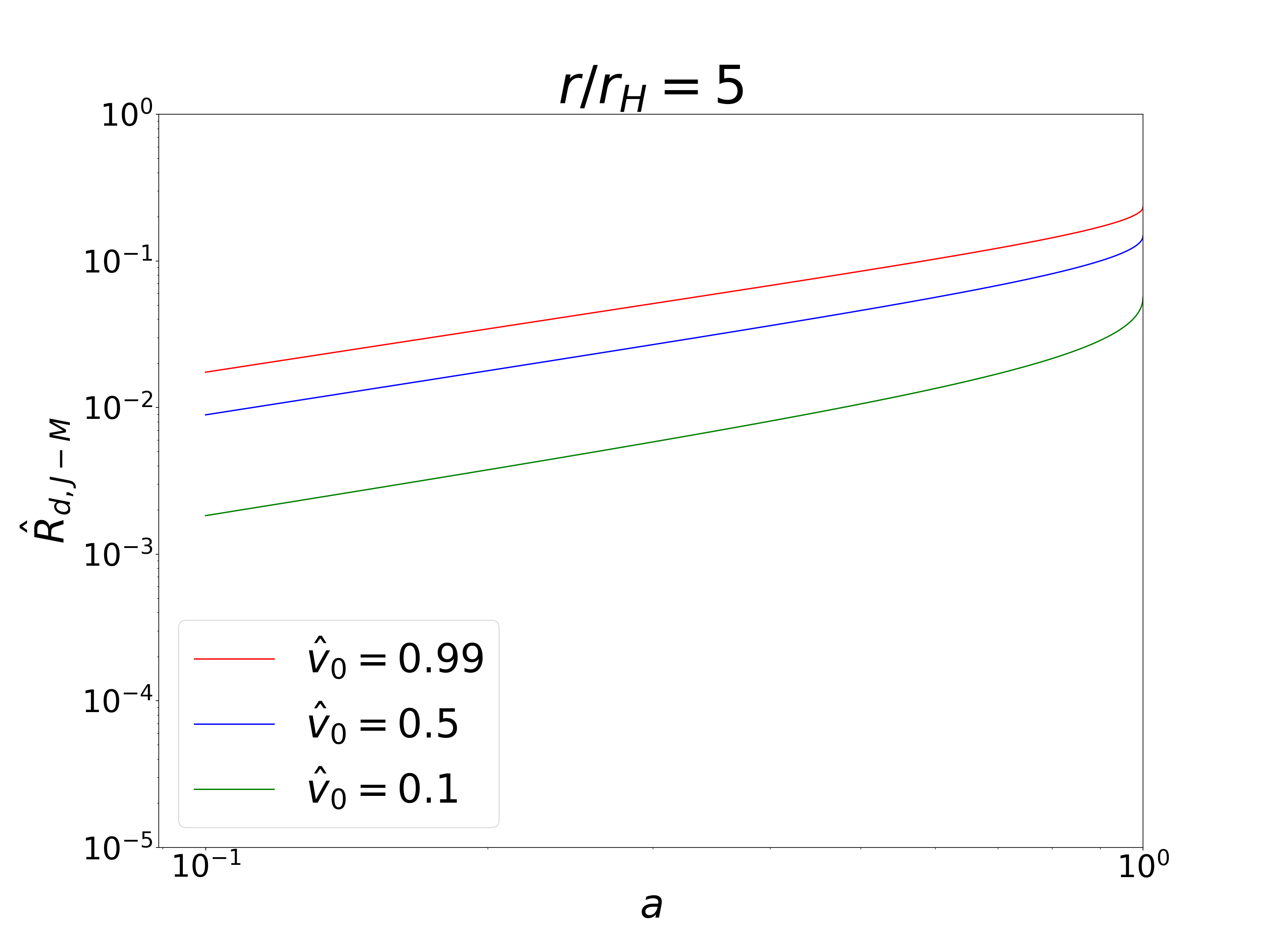}
    \includegraphics[width=0.32\linewidth]{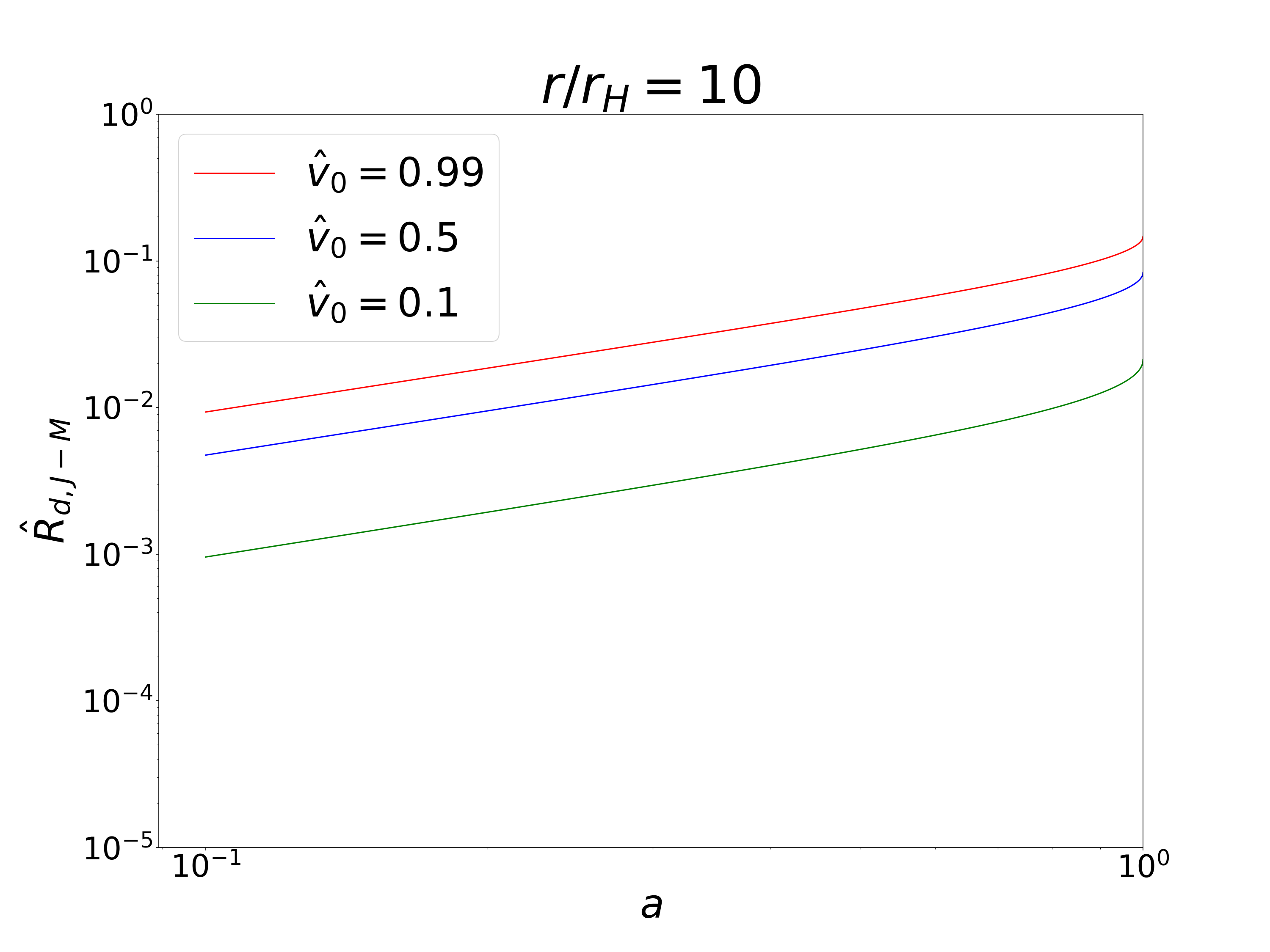}
    \caption{The variations of relative spin-induced-drift rate $\hat{R}_{{\rm d},J-M}$ as functions of radius (top panels), 0th order observed velocity (middle panels) and BH spin (bottom panels) for \textit{case II}. Here $r/r_{\rm H}=1^+$ represents the radius outside but very close to the event horizon.}
    \label{fig:hatRJM}
\end{figure*}

\begin{figure*}
    \centering
    \includegraphics[width=0.32\linewidth]{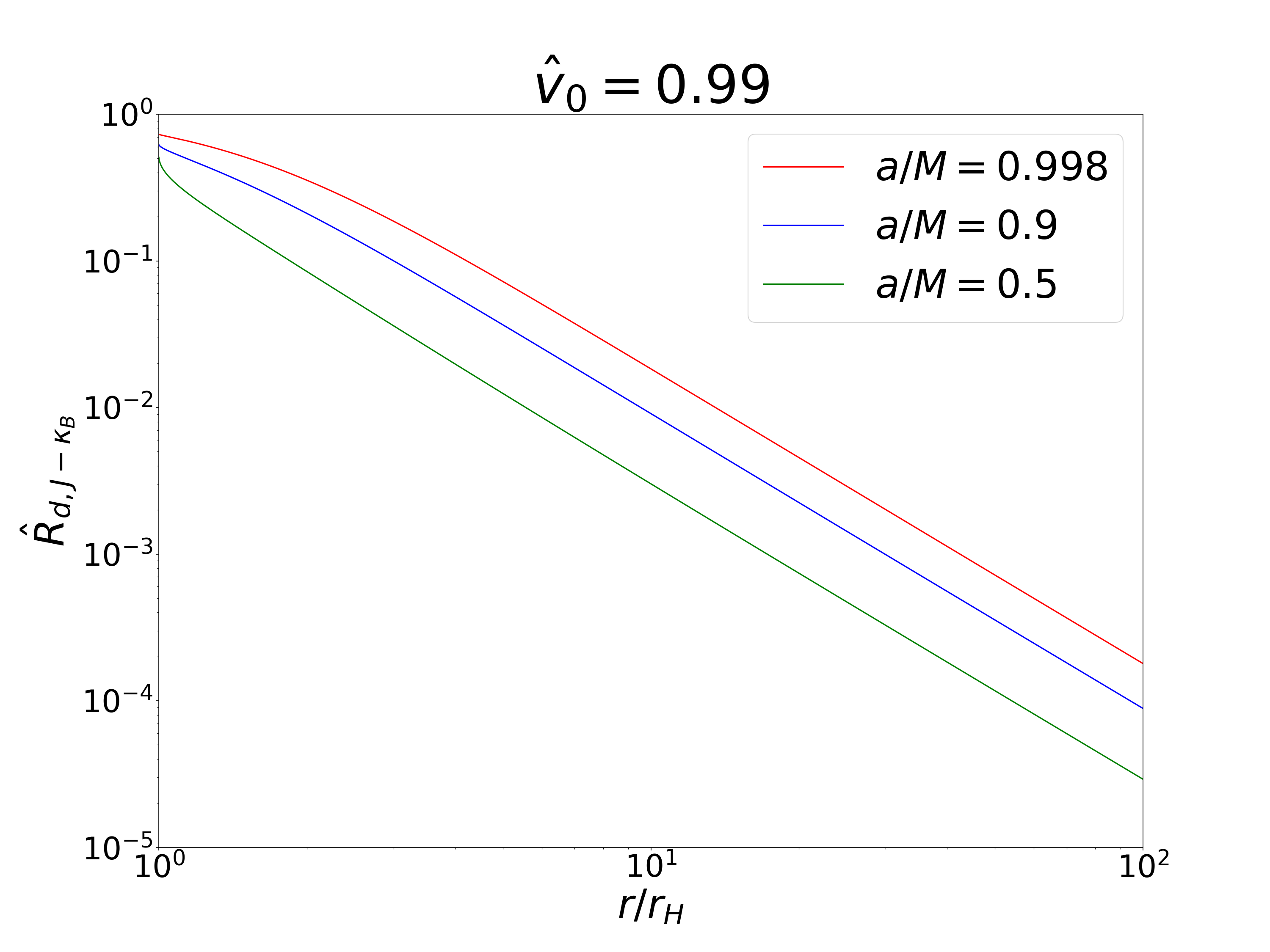}
    \includegraphics[width=0.32\linewidth]{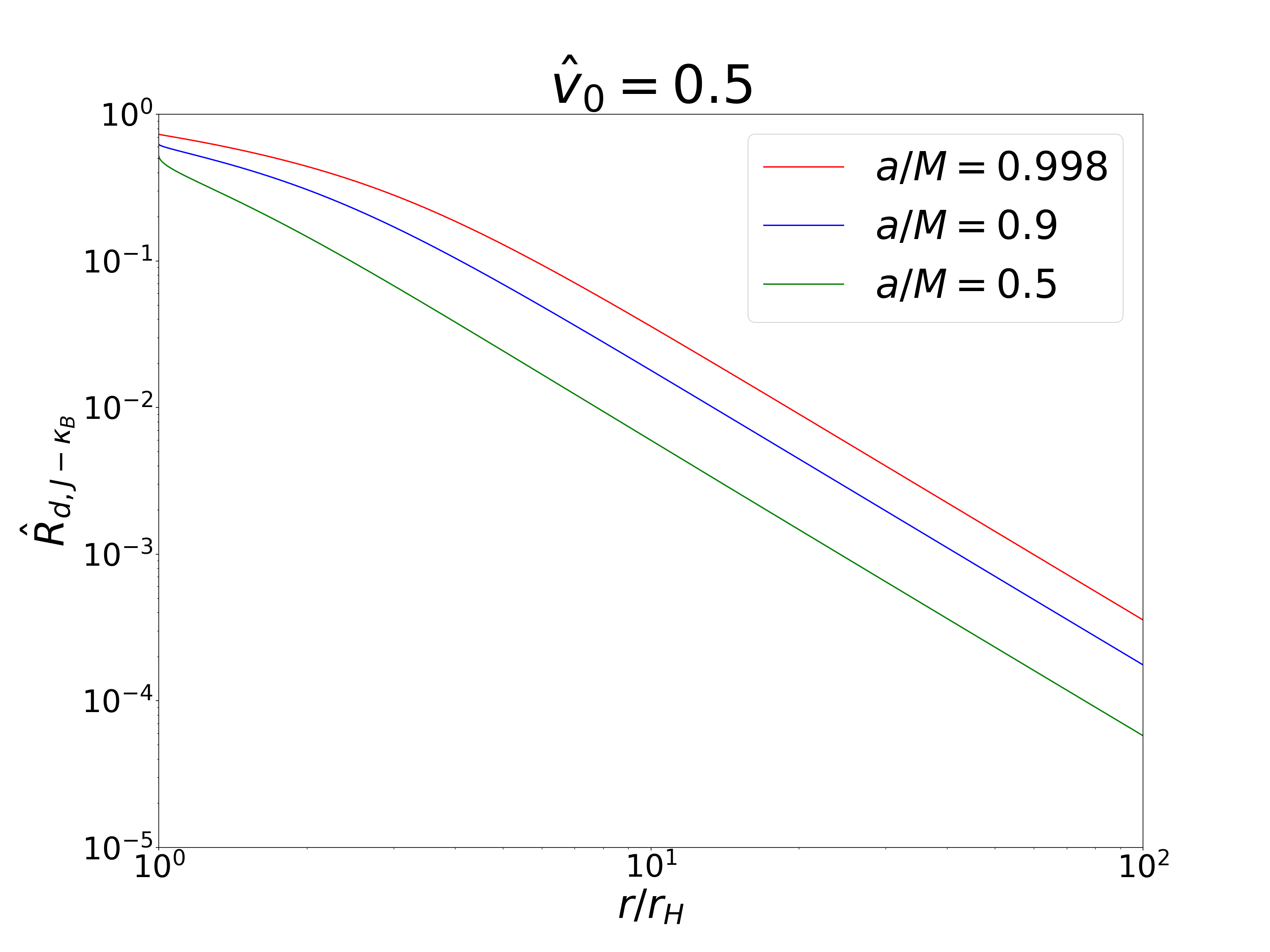}
    \includegraphics[width=0.32\linewidth]{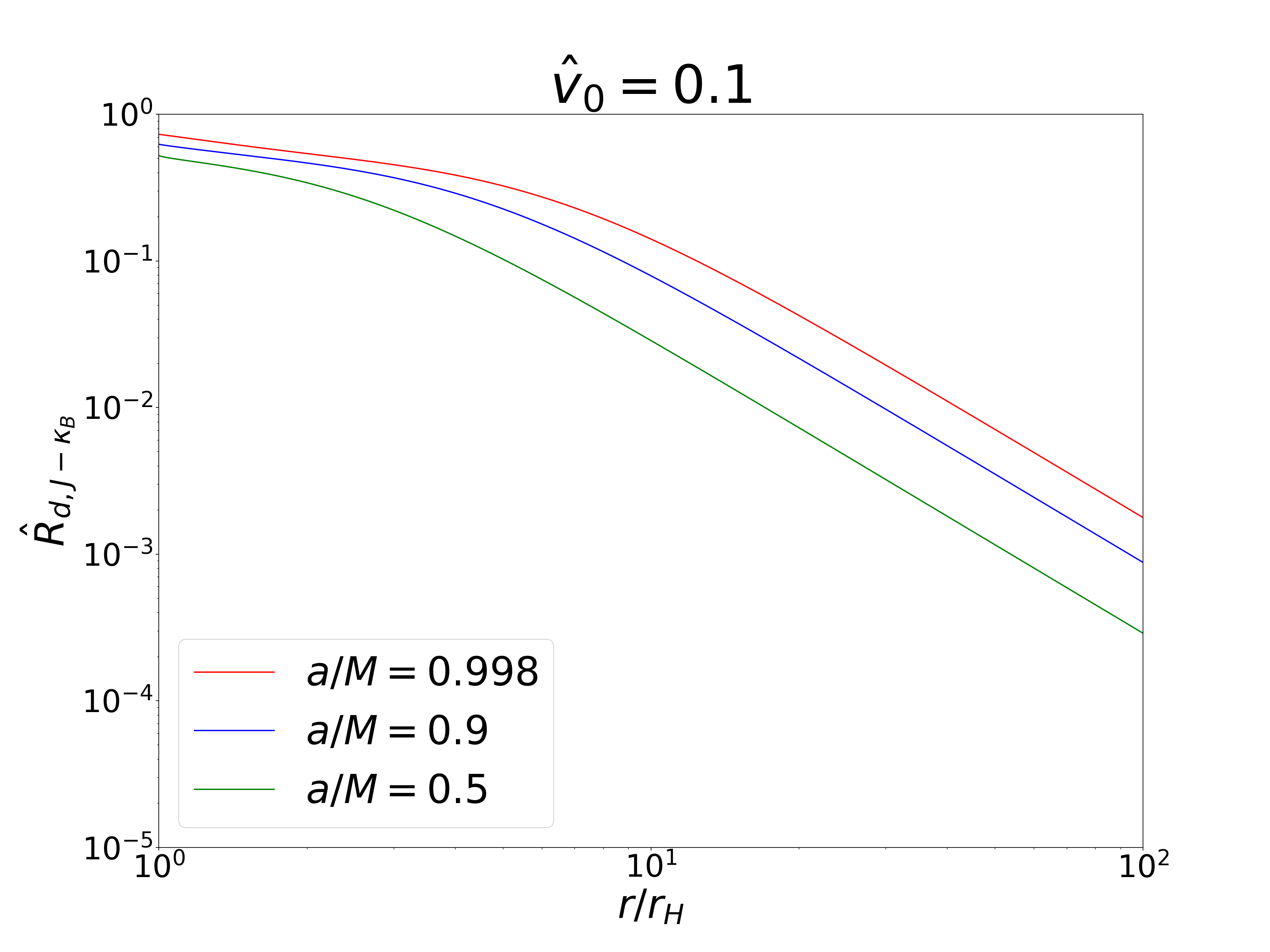}
    \includegraphics[width=0.32\linewidth]{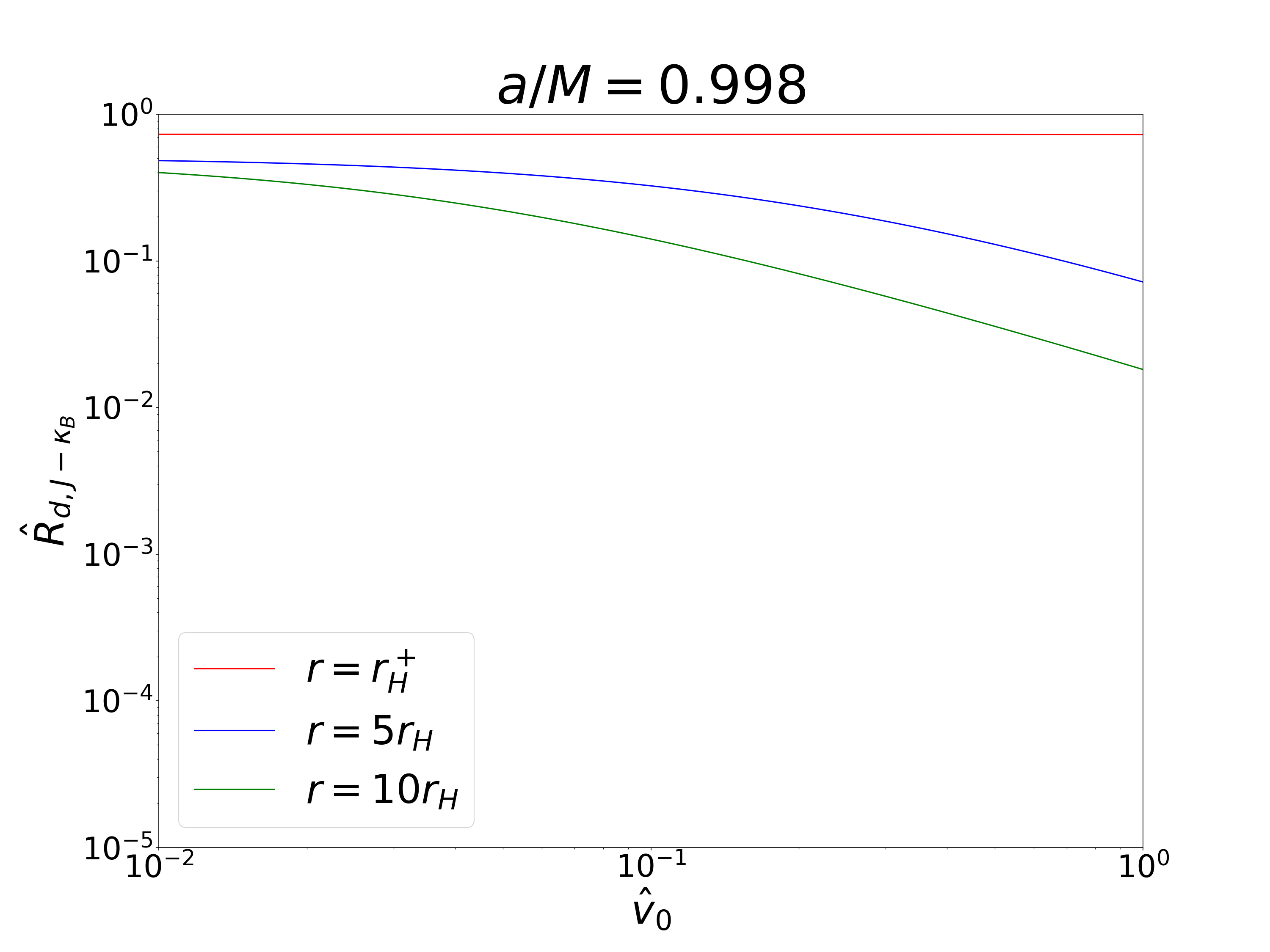}
    \includegraphics[width=0.32\linewidth]{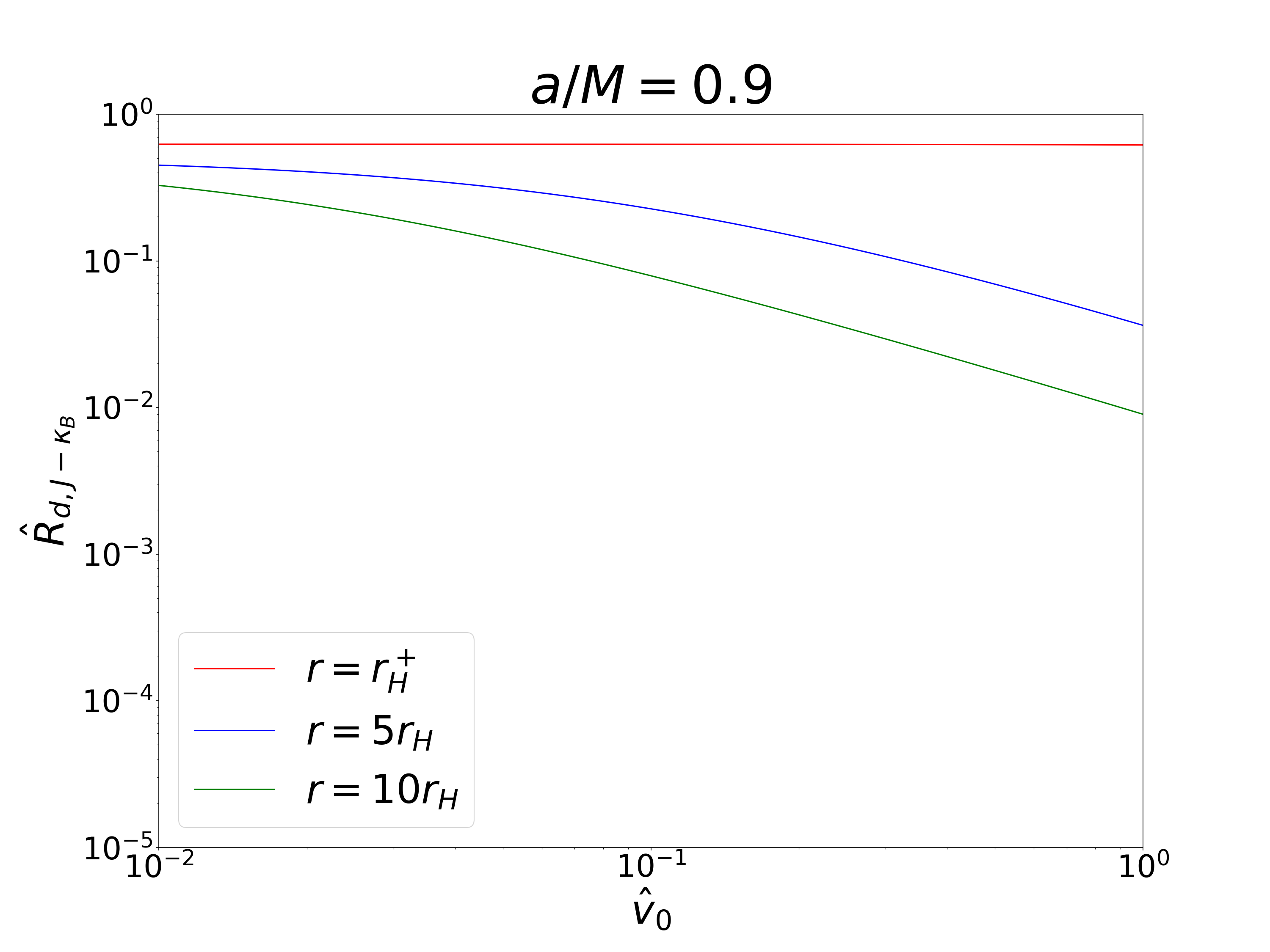}
    \includegraphics[width=0.32\linewidth]{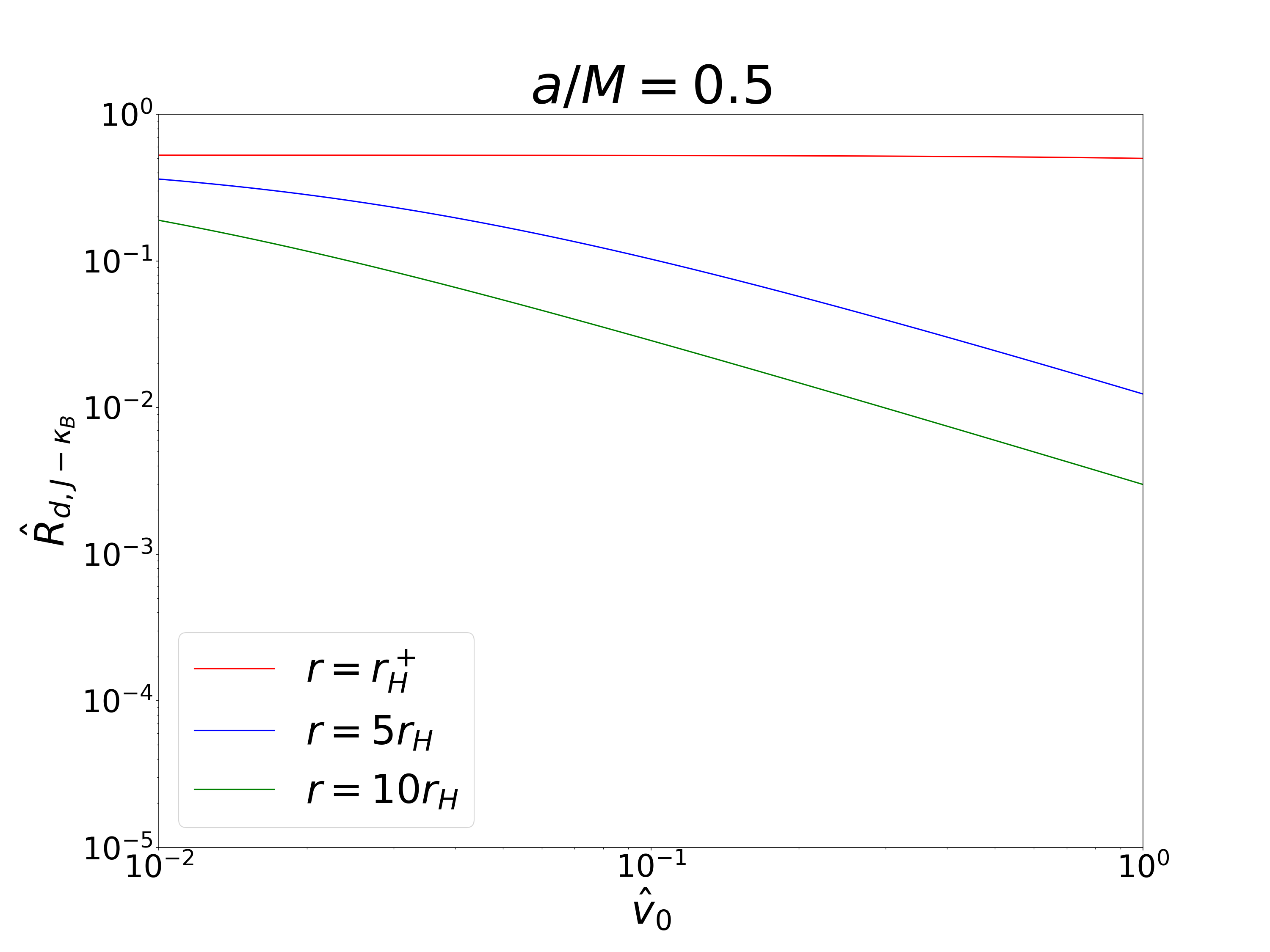}
    \includegraphics[width=0.32\linewidth]{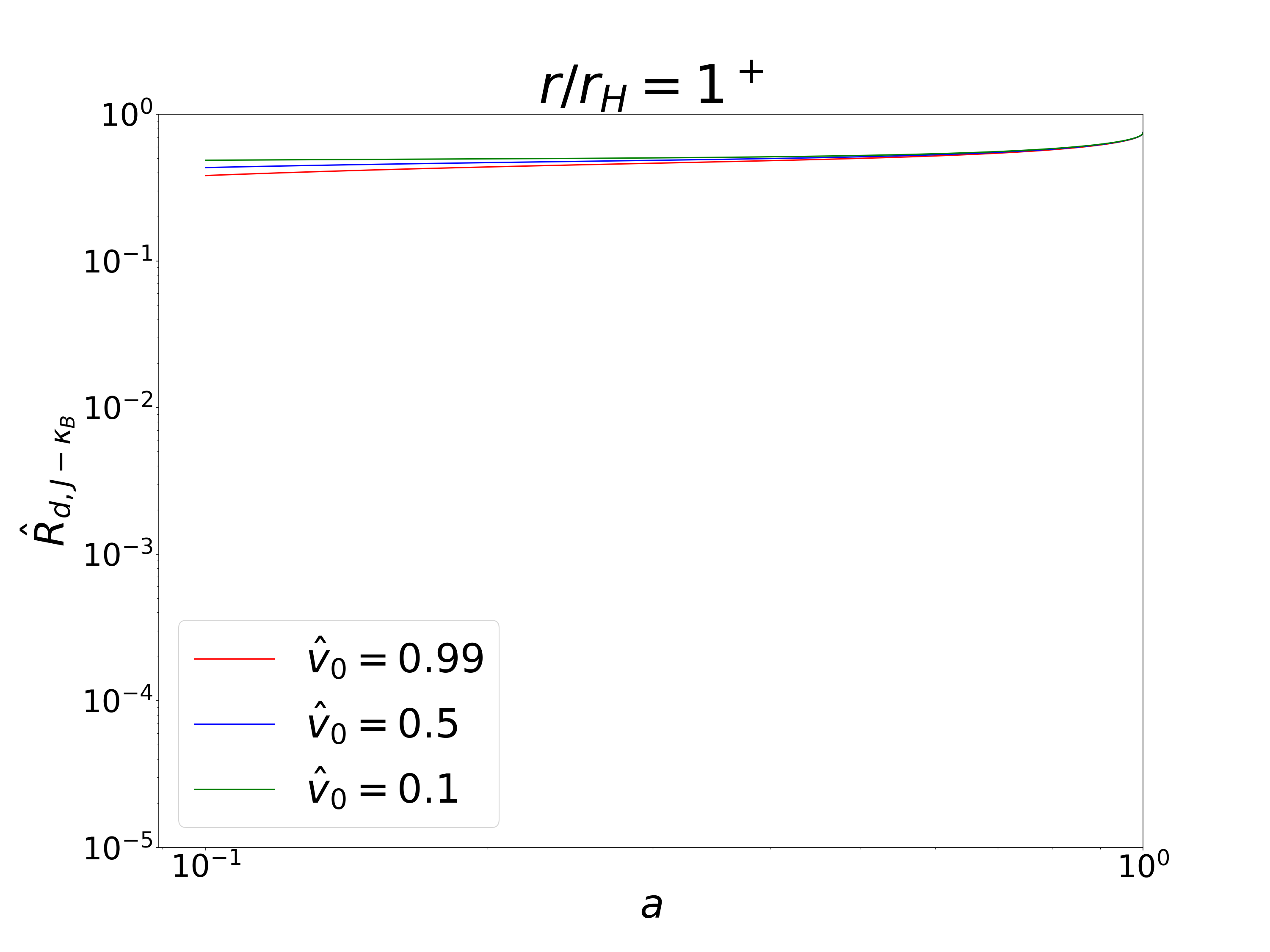}
    \includegraphics[width=0.32\linewidth]{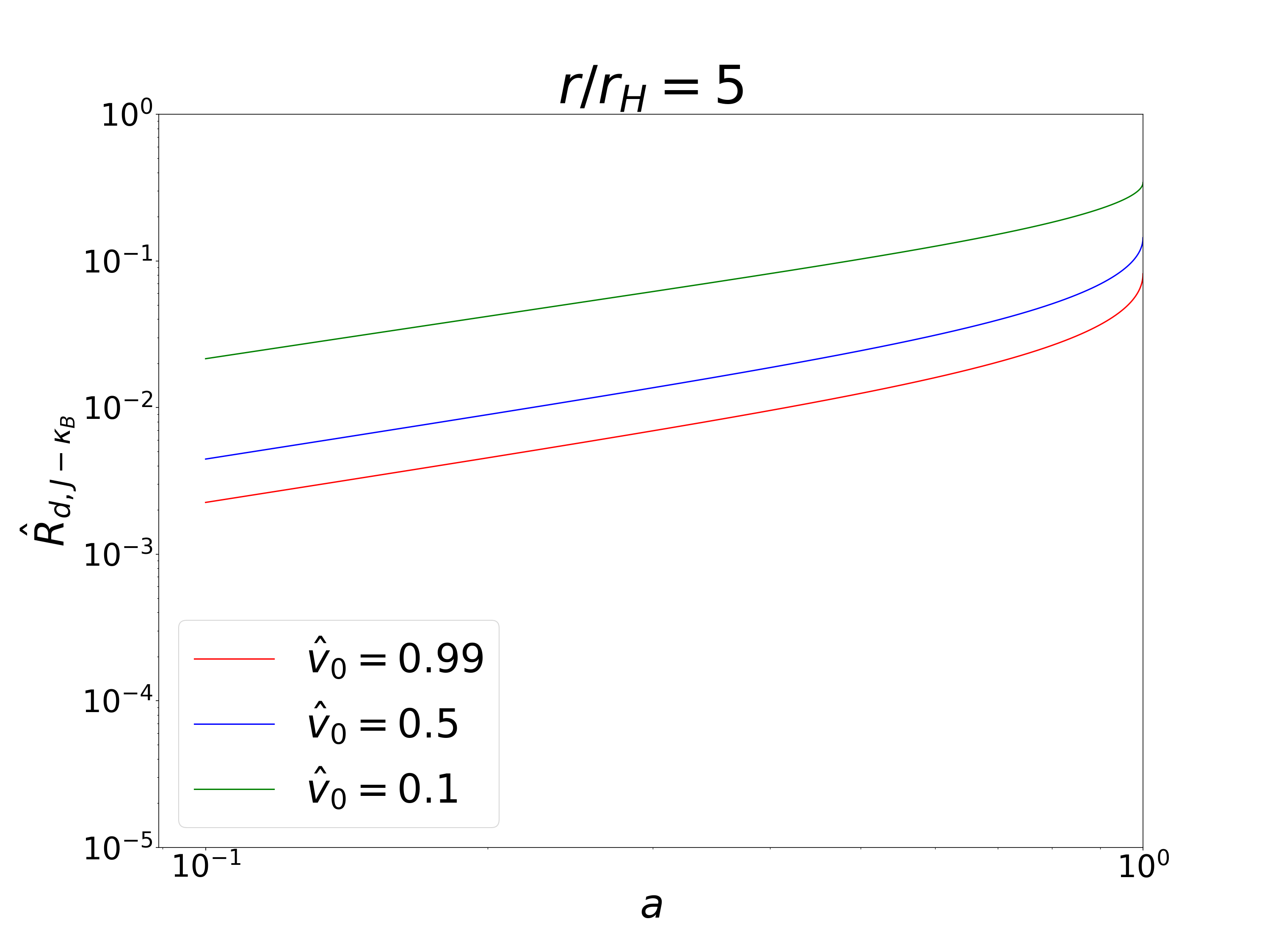}
    \includegraphics[width=0.32\linewidth]{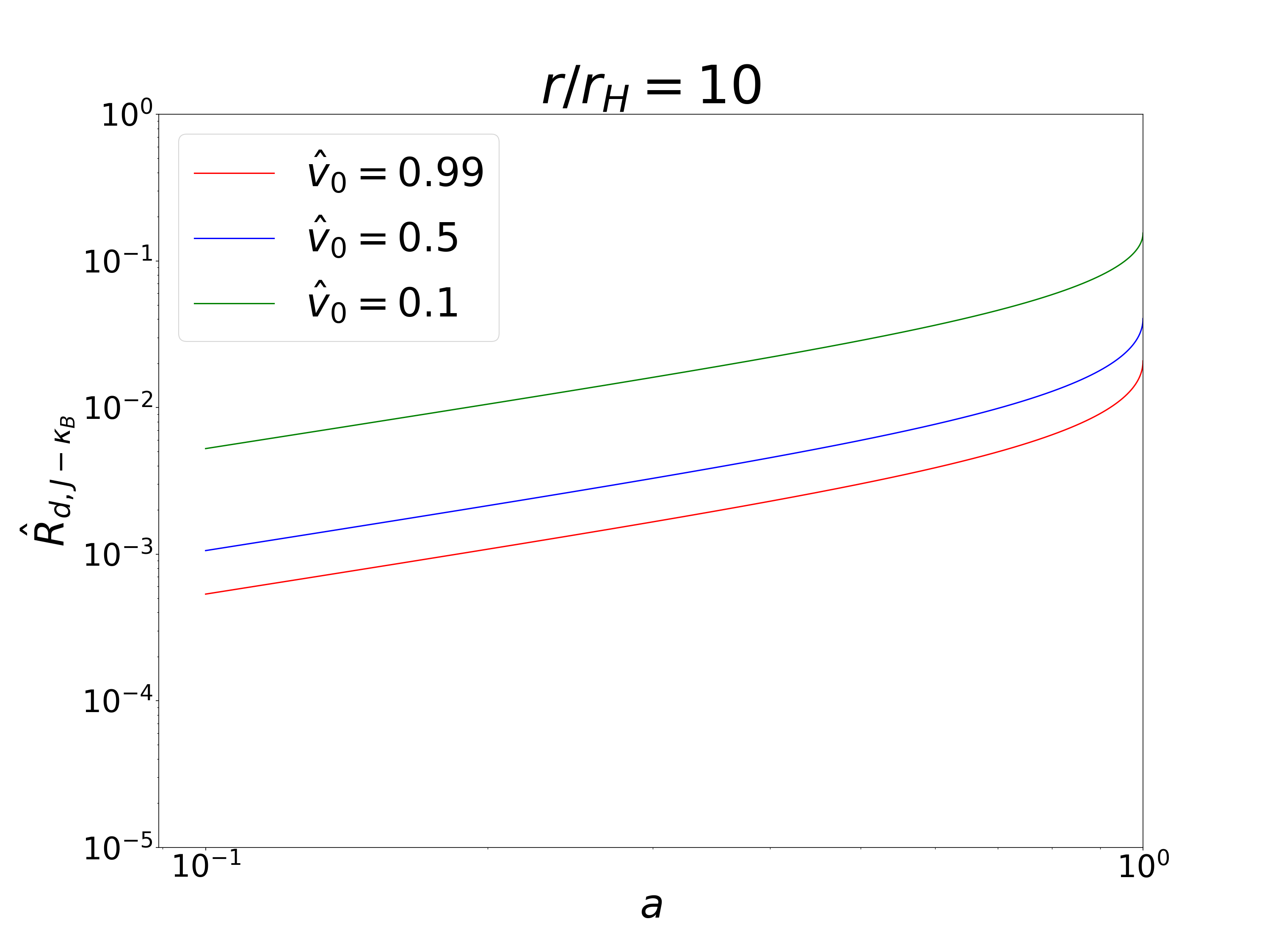}
    \caption{The variations of relative spin-induced-drift rate $\hat{R}_{{\rm d},J-\Vec{\kappa}_B}$ as functions of radius (top panels), 0th order observed velocity (middle panels) and BH spin (bottom panels) for \textit{case II}. Here $r/r_{\rm H}=1^+$ represents the radius outside but very close to the event horizon.}
    \label{fig:hatRJB}
\end{figure*}

In Fig.~\ref{fig:Rd1}, we plot the variations of the drift rate for \textit{case I} (purely radial magnetic field on the equatorial plane). The radius is scaled by the radius of the event horizon $r_{\rm H}$. The drift rate is exhibited in unit of $\frac{m}{eBM}$. The variation with $r$ for fixed $\hat{v}_0$ and $a$, the variation with $\hat{v}_0$ for fixed $r$ and $a$, and the variation with $a$ for fixed $r$ and $\hat{v}_0$ are displayed in the top, middle, and bottom panels, respectively (same for Fig.~\ref{fig:Rd2}-\ref{fig:hatRJB}). All lines terminate at the point where Eq.~\eqref{eq:condition} is violated. From the top panels, the drift rate decays as $r^{-3}$, consistent with the Newtonian result. The sudden drops at small $r$ (for fixed $\hat{v}_0$) correspond to $\hat{v}_{0,r}\rightarrow 0$ caused by effective frame dragging (see Eq.~\eqref{eq:rate-case1}), the same reason accounts for the sudden drops at large $a$ seen in the bottom panels. The middle panels indicate that the drift rate is not sensitive to the variation in $\hat{v}_0$, the observed leading order velocity, except in extreme situations. 

All panels in Fig.~\ref{fig:Rd1} indicate that, for the case of a radial magnetic field on the equatorial plane, the drift induced by BH spin can be considerable. In particular, for $r\lesssim 10r_{\rm H}$, $\hat{v}_0\gtrsim 0.9$ and $a\gtrsim 0.9M$, the drift rate satisfies $R_{\rm d}M^2 \gtrsim (mM)/(eB)$. A significant drift requires $R_{\rm d}\gtrsim 10^{-2}$. Let us consider an electron (or a positron) moving in the magnetosphere of a BH, with $m\approx 9.0\times 10^{-31} {\rm kg}$ and $e\approx 1.6\times 10^{-19} {\rm C}$. We require the magnetic field to be strong enough that the guiding-center approximation is valid. Meanwhile, the magnetic field strength cannot be too high, otherwise the drift becomes insignificant. The validity of the guiding-center approximation and the significance of the drift induced by gravitational effect (due to mass or spin) require $10^{-2}\lesssim R_{\rm d}\lesssim 10^0$. This implies that the BH mass and the magnetic strength near the event horizon should satisfy:
\begin{equation}
    \begin{aligned}
        {\rm for~proton:~~}& 10^{-1}M_{\odot}\cdot{\rm G}\lesssim MB \lesssim 10 M_{\odot}\cdot{\rm G} \\
        {\rm for~electron:~~}& 10^{-4}M_{\odot}\cdot{\rm G}\lesssim MB \lesssim 10^{-2}M_{\odot}\cdot{\rm G}
    \end{aligned}
    \label{eq:magnitude}
\end{equation}
The drift induced by gravity (BH mass) is significant near a BH whose mass and local magnetic field satisfy the condition in Eq.~\eqref{eq:magnitude}. If this BH has a relatively high angular momentum ($a\gtrsim 0.9M$), the spin induced drift also becomes significant near the event horizon.

The variations of the drift rate for \textit{case II} (purely toroidal magnetic field on the equatorial plane) are plotted in Fig.~\ref{fig:Rd2}. Unlike \textit{case I}, three factors now contribute to the drift of a charged particle. Furthermore, the drifts induced by BH spin and magnetic curvature act in the opposite direction to that induced by BH mass. Consequently, as seen from the top and middle panels, there exist critical radii or critical observed leading order velocities that truncate the curves, across which the drift changes the direction. As $r\rightarrow r_{\rm H}$, the contribution of BH mass dominates the drift. In contrast, the magnetic curvature dominates for $r\gg r_{\rm H}$, since the drift induced by magnetic curvature decays as $r^{-1}$, consistent with the Newtonian results. Meanwhile, the contribution of BH mass dominates in the low speed limit, while the contribution of magnetic curvature dominates in the high speed limit. The bottom panels imply that the drift rate is not very sensitive to the BH spin, since the contribution of BH spin is not vitally important compared with those of BH mass and magnetic curvature. It can also be seen from Fig.~\ref{fig:Rd2} that the drift becomes significant when the charged particle approaches the event horizon in the case of purely toroidal magnetic field. The drift rate could reach magnitudes of $R_{\rm d}M^2 \gtrsim (mM)/(eB)$, similar to those in \textit{case I}.

The relative spin-induced drift rates $\hat{R}_{{\rm d},J-M}$ and $\hat{R}_{{\rm d},J-\Vec{\kappa}_B}$ are shown in Fig.~\ref{fig:hatRJM} and Fig.~\ref{fig:hatRJB}, respectively, in order to assess the importance of BH spin for the drift of a charged particle in different situations. It is evident that, in the small-radius and high-spin limit, the contribution of BH spin to the drift is nearly as important as those of BH mass and magnetic curvature. In the limit of large leading order velocity, the contribution of BH spin generally becomes comparable to that of BH mass. This could be anticipated from Eq.~\eqref{eq:rate-case2}, as the drift induced by BH mass in \textit{case II} actually weakens with increasing leading order velocity. When compared with the contribution of magnetic curvature, the BH spin is worth considering generally in the limit of low leading order velocity. Furthermore, when the charged particle is extremely close to the event horizon, the relative spin-induced drift rate becomes insensitive to the leading order velocity, as shown in the middle panels of Fig.~\ref{fig:hatRJM} and Fig.~\ref{fig:hatRJB}. As long as $a\gtrsim 0.9M$, the BH spin remains equally important for the drift of a charged particle compared with the BH mass and magnetic curvature, regardless of how fast the guiding-center moves along the magnetic field line. In summary, for the case of a toroidal magnetic field on the equatorial plane of a rapidly spinning BH, the drift induced by BH spin is considerable when the charged particle approaches the event horizon. In less extreme situations, e.g., $a\lesssim 0.9M$ or $r\gtrsim 5r_{\rm H}$, the contribution of BH spin is worth considering compared with that of BH mass but negligible compared with that of magnetic curvature when the charged particle moves extremely fast along the magnetic field line. Conversely, it is worth considering compared with magnetic curvature but trivial compared with BH mass when the charged particle moves extremely slowly along the magnetic field line.

\section{Further Discussions}
\label{sec:discuss}

It is worth mentioning that the gyration of the charged particle does not appear in the EoM shown in Eq.~\eqref{eq:gca-GR2}. Classically, the gyrovelocity satisfies $\left|\Vec{v}_g\right|=\frac{eBR}{m}$, where the gyroradius $R$ is taken to be infinitesimal in the guiding-center approximation. Therefore, the power series expansion of $u^{\mu}$ in orders of $(m/e)^n$ can never capture $\Vec{v}_g$, as no information about the gyroradius is contained. Moreover, it is expected that the gyration of a charged particle shrinks to a point quickly in a strong magnetic field due to efficient synchrotron radiation \cite{Landau:1975pou,Cai:2022mkw}. The absence of gyration in the EoM is thus not unnatural. However, as shown in Eq.~\eqref{eq:vd-gradient}, the drift induced by $\nabla B$ is absent because of the absence of gyration, which is important in classical guiding-center approximation. It is an imperfection of the covariant fomulation shown in Sect.~\ref{sec:gca-GR}. From the Newtonian scheme, it is estimated that $\Vec{v}_{\nabla B}$ vanishes in \textit{case I} (since $\hat{B}\parallel \nabla B$) but exists in \textit{case II}. How to incorporate $\Vec{v}_{\nabla B}$ naturally requires further discussion.

There is no concise expression for $u_{1,\parallel}^{\mu}$, the 1st order 4-velocity of the charged particle stuck on the magnetic sector. One has to obtain $u_{1,\parallel}^{\mu}$ via the 2nd order EoM in Eq.~\eqref{eq:GR-eom2}. Contracting $\epsilon_{\mu\nu}$ on both sides of the 2nd order EoM yields:
\begin{equation}
    0=\epsilon_{\mu\nu}\left(u_0^{\kappa}\nabla_{\kappa}u_1^{\nu}+u_1^{\kappa}\nabla_{\kappa}u_0^{\nu}-\frac{1}{m}f_2^{\nu}\right)
    \label{eq:u1-para}
\end{equation}
where the term with $u_2^{\mu}$ vanishes based on Eq.~\eqref{eq:epsilon}. Eq.~\eqref{eq:u1-para} lies on the 2D field sheet, hence it contains at most two linearly independent differential equations. One could obtain $u_{1,\parallel}^{\mu}$ by integrating $\nabla_{\kappa}u_0^{\nu}$, $\nabla_{\kappa}u_{1,\perp}^{\nu}$ and $f_2^{\nu}$ on the field sheet based on Eq.~\eqref{eq:u1-para}.

\section{Summary and Outlook}
\label{sec:sum}

In this work, we study the drift of a charged particle in the magnetosphere of a BH under both Newtonian and GR schemes. Moreover, we analyze the effect of BH spin and assess whether it is important for the drift of a charged particle. 

We first introduce the guiding-center approximation and the drift of a charged particle in the Newtonian scheme. The drift induced by the rotation of a massive object is derived, though it is negligible compared with the gravity-induced drift in the Newtonian limit. Then we present the covariant formulation of the guiding-center approximation by expanding the 4-velocity of a charged particle as a power series in orders of $(m/e)^n$. We confirm that the GR results reduce to the Newtonian results in the weak gravitational field and low speed limits. We find that the electric field induced drift arises from the 0th order EoM, while the drift induced by gravity and the rotation of a massive object arises from the 1st order EoM. 

We concentrate our discussions on Kerr spacetime and consider two specific cases: the purely radial and purely toroidal magnetic fields lying on the equatorial plane. We define the drift rate, which quantifies the distance a charged particle deviates from the magnetic field line due to drift after traveling a unit distance along the magnetic field. The value of the drift rate indicates the significance of the drift. Our main conclusions, shown in Fig.~\ref{fig:Rd1}-\ref{fig:hatRJB}, are:
\begin{itemize}
    \item For a purely radial magnetic field on the equatorial plane, the BH spin is the only factor causing the drift. This spin-induced drift is worth considering only when a charged particle moves fast along the magnetic field ($\hat{v}_0\gtrsim 0.9$) near a rapidly spinning BH ($r\lesssim 10r_{\rm H}$ and $a\gtrsim 0.9M$).
    \item For a purely toroidal magnetic field, there are three factors--BH mass, BH spin and curvature of magnetic field line--inducing the drift, among which the BH spin plays a relatively minor role. But the spin-induced drift becomes non-negligible, compared with other two factors, when a charged particle is very close to a rapidly spinning BH ($r\lesssim 5r_{\rm H}$ and $a\gtrsim 0.9M$).
    \item When a BH magnetosphere satisfies $10^{-1}M_{\odot}\cdot{\rm G}\lesssim MB \lesssim 10 M_{\odot}\cdot{\rm G}$ for protons or $10^{-4}M_{\odot}\cdot{\rm G}\lesssim MB \lesssim 10^{-2} M_{\odot}\cdot{\rm G}$ for electrons, the gravity induced drift becomes important. If this BH has a relatively high angular momentum, the spin induced drift also becomes important near the event horizon.
\end{itemize}

This work is a preliminary study, providing a glimpse on how a charged particle drift in a BH magnetosphere under the induction of BH spin. It shows when the BH spin, which is absent in the Newtonian or special relativistic scheme, is worth considering compared with other factors when we discuss the drifts of charged particles in a BH magnetosphere. In this work, we only discuss two ideal but typical cases, a purely radial and toroidal magnetic field on the equatorial plane. On the one hand, it is supported by the theory of FFE and PIC simulations that the equatorial current sheet forms near the BH in highly magnetized accretion flow \cite{Camilloni:2022kmx,Gelles:2024tpz,Vos:2024loa}, implying that the magnetic field is mainly composed of radial and toroidal components near the equatorial plane. On the other hand, although the magnetic field is believed to be highly turbulent in a real accretion system after a long time evolution \cite{Nathanail:2020wap,Ripperda:2020bpz,Ripperda:2021zpn}, the impact of BH spin on the drift in the general case should be approximately a combination of those in \textit{case I} and \textit{case II}.

This work has several limitations that require further exploration. First, as mentioned in Sect.~\ref{sec:discuss}, the absence of gyration in the EoM destines the absence of drift induced by $\nabla B$, which is not negligible in the guiding-center approximation classically. Second, unfortunately, it is difficult to observe a BH magnetosphere that satisfies Eq.~\eqref{eq:magnitude} and in which the spin-induced drift is significant. The low BH mass and weak magnetic strength (and therefore the low accretion rate) make such accretion system less luminous. A0620-00 is a promising source, with $M\simeq M_{\odot}$ \cite{A0620-00-mass} and $B\simeq 1{\rm G}$ \cite{Kravtsov:2024ovq}. However, its BH spin was believed to be low ($a\simeq 0.12$) \cite{A0620-00-spin}, so that the spin induced drift remains negligible. Another promising source is V404~Cygni, with $M\simeq 10M_{\odot}$ \cite{V404Cygni-mass} and $a\gtrsim 0.9$ \cite{Walton:2016fso}. It has been suggested that the magnetic field strength of V404~Cygni is approximately $0.5{\rm G}$ \cite{Chandra:2017fhp}, with which the drift of proton near the event horizon would be significant. All in all, we hope that advances in observational techniques will help us identify more BH accretion systems in which the spin induced drift is significant. Third, in general, charged particles move in the BH magnetosphere in populations, constituting a sparse accretion flow. Whether the drift influences the behavior of the accretion flow and how important BH spin could be are worthy of further investigation.

\begin{acknowledgments}
\begin{CJK}{UTF8}{gbsn}
I am grateful for the support of my parents. I would like to thank Dr. Yuedan Wang (汪月丹) for her consistent companionship. I appreciate the constructive suggestions given by the anonymous referee. This work is supported by NSFC (Grant No.~12605090).
\end{CJK}
\end{acknowledgments}

\appendix

\section{Christoffel Symbols}
\label{sec:Christoffel}

To help the readers follow our calculations, here we list the non-zero components of Christoffel symbols of Kerr metric in BL coordinates when $\theta=\pi/2$ and their approximations in the large radius limit:
\begin{equation}
    \Gamma^{t}_{tr}=\Gamma^{t}_{rt}=~
    \frac{M\left(r^2+a^2\right)}{r^2\Delta} 
    ~~~\xrightarrow{r~\rightarrow~\infty}~~~ \frac{M}{r^2}
\end{equation}
\begin{equation}
    \Gamma^{t}_{r\phi}=\Gamma^{t}_{\phi r}=~
    -\frac{Ma\left(r^2+a^2\right)}{r^2\Delta}
    ~~~\xrightarrow{r~\rightarrow~\infty}~~~ -\frac{J}{r^2}
\end{equation}
\begin{equation}
    \Gamma^{r}_{tt}=~
    \frac{M\Delta}{r^4}
    ~~~\xrightarrow{r~\rightarrow~\infty}~~~ \frac{M}{r^2}
\end{equation}
\begin{equation}
    \Gamma^{r}_{t\phi}=\Gamma^{r}_{\phi t}=~
    -\frac{Ma\Delta}{r^4}
    ~~~\xrightarrow{r~\rightarrow~\infty}~~~ -\frac{J}{r^2}
\end{equation}
\begin{equation}
    \Gamma^r_{rr}=~
    \frac{a^2-Mr}{r\Delta}
    ~~~\xrightarrow{r~\rightarrow~\infty}~~~ -\frac{M}{r^2}
\end{equation}
\begin{equation}
    \Gamma^{r}_{\theta\theta}=~
    \frac{\Delta}{r}
    ~~~\xrightarrow{r~\rightarrow~\infty}~~~ r
\end{equation}
\begin{equation}
    \Gamma^{r}_{\phi\phi}=~
    \frac{\left(Ma^2-r^3\right)\Delta}{r^4}
    ~~~\xrightarrow{r~\rightarrow~\infty}~~~ -r
\end{equation}
\begin{equation}
    \Gamma^{\theta}_{r\theta}=\Gamma^{\theta}_{\theta r}=~
    \frac{1}{r}
\end{equation}
\begin{equation}
    \Gamma^{\phi}_{tr}=\Gamma^{\phi}_{rt}=~
    \frac{Ma}{r^2\Delta}
    ~~~\xrightarrow{r~\rightarrow~\infty}~~~ \frac{J}{r^4}
\end{equation}
\begin{equation}
    \Gamma^{\phi}_{r\phi}=\Gamma^{\phi}_{\phi r}=~
    \frac{r^3-2Mr^2-Ma^2}{r^2\Delta}
    ~~~\xrightarrow{r~\rightarrow~\infty}~~~ \frac{1}{r}
\end{equation}

\nocite{*}

\bibliography{refs}

\end{document}